\documentstyle[12pt,epsf]{article}
\def \sect #1 {\setcounter{equation} 0\section{#1}}

\def \be  {\begin{equation}}
\def \ee  {\end{equation}}
\def \ba  {\begin{eqnarray}}
\def \ea  {\end{eqnarray}}
\def \baa {\begin{eqnarray*}}
\def \eaa {\end{eqnarray*}}
\def \bb  {\begin{thebibliography}}
\def \eb  {\end{thebibliography}}
\newcommand \ci [1] {\cite{#1}}
\newcommand \bi [1] {\bibitem{#1}}
\def \lab #1 {\label{#1}}
\newcommand\re[1]{(\ref{#1})}

\newcommand\lr[1]{{\left({#1}\right)}}

\def \tr {\mbox{tr\,}}
\newcommand \vev [1] {\langle{#1}\rangle}

\newcommand \ket [1] {|{#1}\rangle}

\def \e {\mbox{e}}
\def \CO {{\cal O}}

\def \CM {{\cal M}}

\def \CJ {{\cal J}}
\def \res {\mathop{\rm res}\nolimits}
\newcommand \widebar [1] {\overline{#1}}

\font\cmss=cmss10 \font\cmsss=cmss10 at 11pt
\def\inbar{\,\vrule height1.5ex width.4pt depth0pt}
\def\IC{\relax\hbox{$\inbar\kern-.3em{\rm C}$}}
\def\IZ{\relax\ifmmode\mathchoice
{\hbox{\cmss Z\kern-.4em Z}}{\hbox{\cmss Z\kern-.4em Z}}
{\lower.9pt\hbox{\cmsss Z\kern-.4em Z}}
{\lower1.2pt\hbox{\cmsss Z\kern-.4em Z}}\else{\cmss Z\kern-.4em Z}\fi}
\def\IR{{\hbox{{\rm I}\kern-.2em\hbox{\rm R}}}}

\def\IP{{\hbox{{\rm I}\kern-.2em\hbox{\rm P}}}}
\def\II{\hbox{{1}\kern-.25em\hbox{l}}}

\def\Im{\hbox{\rm Im}\,}
\newcommand{\as}{\ifmmode\alpha_{\rm s}\else{$\alpha_{\rm s}$}\fi}

\begin{document}

\def\thefootnote{\fnsymbol{footnote}}
\thispagestyle{empty}
\hfill\parbox{50mm}{{\sc LPTHE--Orsay--96--76} \par
                         hep-th/9609123  \par
                         September, 1996}
\vspace*{35mm}
\begin{center}
{\LARGE Integrable structures and duality in high-energy QCD}
\par\vspace*{15mm}\par
{\large G.~P.~Korchemsky}%
\footnote{E-mail: korchems@qcd.th.u-psud.fr}
\par\bigskip\par\medskip

{\em Laboratoire de Physique Th\'eorique et Hautes Energies%
\footnote{Laboratoire associ\'e au Centre National de la Recherche
Scientifique (URA D063)}
\\
Universit\'e de Paris XI, Centre d'Orsay, b\^at. 211\\
91405 Orsay C\'edex, France
\\[3mm]
and
\\[3mm]
Laboratory of Theoretical Physics
\\
Joint Institute for Nuclear Research
\\
141980 Dubna, Russia}

\end{center}
\vspace*{12mm}

\begin{abstract}
We study the properties of color-singlet Reggeon compound states in
multicolor high-energy QCD in four dimensions. Their spectrum is governed
by completely integrable (1+1)-dimensional effective QCD Hamiltonian whose
diagonalization within the Bethe Ansatz leads to the Baxter equation for
the Heisenberg spin magnet. We show that nonlinear WKB solution of the
Baxter equation gives rise to the same integrable structures as appeared
in the Seiberg-Witten solution for $N=2$ SUSY QCD and in the finite-gap
solutions of the soliton equations. We explain the origin of hyperelliptic
Riemann surfaces out of QCD in the Regge limit and discuss the meaning of
the Whitham dynamics on the moduli space of quantum numbers of the Reggeon
compound states, QCD Pomerons and Odderons.
\end{abstract}

\vspace*{15mm}
\begin{center}
{\sl Dedicated to the memory of Victor~I.~Ogievetsky}
\end{center}

\newpage
\def\thefootnote{\arabic{footnote}}
\setcounter{footnote} 0

\sect{Introduction}

In the last decades a tremendous progress has been achieved in
understanding of low-dimensional integrable field theories and
statistical models. Apart from the fact that their study serves as
a laboratory for testing a new ideas, it has been expected for a
long time that low-dimensional integrable systems should appear in
the description of the effective dynamics of quantized (3+1)--dimensional
Yang-Mills theories (see e.g. \ci{book}). These expectations were 
confirmed in 1994 by two examples coming from different problems: the exact
calculation of low-energy effective action of $N=2$ supersymmetric
Yang-Mills theory \ci{SW} and study of the Regge asymptotics of
perturbative QCD \ci{FK,Lip}. It was found that well-known
integrable structures emerge in both problems.

In the first example \ci{SW}, the Seiberg--Witten solution for the
low-energy effective action of pure $N=2$ SUSY Yang-Mills theory with
the $SU(2)$ gauge group and its generalization to other gauge
groups \ci{gen} were interpreted within the framework of finite-gap
solutions of the soliton equations and their Whitham deformations
\ci{GKM3}. In particular, the moduli space of vacua of $N=2$ SUSY
$SU(N_c)$ Yang-Mills theory without matter was identified as a moduli
space of hyperelliptic curve $\Gamma$, which amazingly enough coincides with
the spectral curve of 1-dimensional periodic Toda chain with $N_c$
sites \ci{GKM3,To}. Moreover, including the matter hypermultiplets in the
adjoint and fundamental representations leads to a deformation of
the Toda spectral curve into the spectral curves corresponding to
the Calogero-Moser integrable models \ci{CM,IM} and Heisenberg spin
chains \ci{spin}, respectively. The effective energy and the BPS
spectrum of excitations can be expressed in terms of periods of a
certain meromorphic 1-differential on $\Gamma$, which has
a defining property that its external derivative is a holomorphic
differential on $\Gamma$. This differential was
identified as a generating functional of the Whitham equations
describing the adiabatic deformations of soliton solutions \ci{Wh}.
There were different proposals to interpret the Whitham dynamics
\ci{GKM3,IM,TN,G} but, despite of a lot of efforts, the connection
between integrable systems and the Seiberg-Witten solution still has a
status of an observation and the underlying dynamical mechanism
remains mysterious.

Our second example is related to the asymptotics of QCD at high energy
and fixed transferred momentum, $s \gg -t$, the famous Regge limit \ci{Col}.
It has been observed many years ago (see e.g. \ci{CW}) that perturbative
expressions for the scattering amplitudes in QCD in the Regge limit have a
striking similarity with predictions of a low-dimensional field theory.
Indeed, it was realized that in the Regge limit QCD should
be replaced by an effective $(2+1)-$dimensional Reggeon field theory
\ci{Gr}, which should inherit all (still unknown) quantum symmetries of QCD.
Recently, different attempts have been undertaken to deduce its effective
action out of QCD \ci{Leff,VV}. The resulting expressions become very
complicated and the exact solution of the $S-$matrix of the Reggeon effective
theory remains problematic. Nevertheless, there exists a meaningful
approximation to the Reggeon effective theory, the so-called generalized
leading logarithmic approximation (LLA) \ci{BKP}, in which the $S-$matrix
of QCD exhibits remarkable properties of integrability. It turns out that
in this approximation QCD is effectively described by the 1-dimensional
{\it quantum\/} XXX Heisenberg magnet of spin $s=0$ \ci{FK,Lip}. As a
result, one can calculate the Regge asymptotics of the scattering
amplitudes by applying a powerful Quantum Inverse Scattering Method 
\ci{QISM}. In particular, the Bethe Ansatz solution of perturbative 
QCD $S-$matrix in the generalized LLA has been developed in \ci{FK,Bet}.
It is based on the Separation of Variables and leads to the  
Baxter equation for the XXX Heisenberg magnet \ci{FM,Guz,SoV}.

The exact solution of the Baxter equation was established in a few
special cases \ci{FK,Bet} and different approaches to its general solution
were proposed in \ci{MW,Qua,J,JW}. One of them is based on the quasiclassical
expansion of the Baxter equation \ci{Qua}. It allows to obtain asymptotic
approximations to the solutions of the Baxter equation, which are in a good
agreement with the results of numerical calculations.

In the present paper we continue the study of the Baxter equation started
in \ci{Qua}. We show that the WKB solution of the Baxter equation,
describing the Regge limit of QCD, gives rise to the same integrable 
structures as appeared in the Seiberg-Witten solution for $N=2$ SUSY QCD
and in the finite-gap solutions of the soliton equations. We explain the
origin of hyperelliptic Riemann surfaces out of QCD in the Regge limit
and discuss the meaning of Whitham dynamics on their moduli space.

The paper is organized as follows. In Sect.~2 we demonstrate the relation
between high-energy QCD in the Regge limit and XXX Heisenberg magnet.
The hamiltonian of the 1-dimensional XXX Heisenberg magnet appears as
a kernel of the Bethe-Salpeter equation for the partial waves of the 
scattering amplitudes in multi-color QCD. Its diagonalization is
performed in Sect.~3 by applying the method of Separation of Variables 
\ci{FM,Guz,SoV}, which leads to the Baxter equation. In Sect.~4 we
develop a nonlinear WKB expansion of the Baxter equation. We describe in
detail the classical mechanics induced by the leading term of the WKB 
solution and establish a close relation between Reggeon compound states 
in QCD and soliton solutions of the KP/Toda hierarchy \ci{NMPZ,Kr}. The 
quantization conditions for the Reggeon compound states and their 
interpretation in terms of the Whitham flows are discussed in Sect.~5. The 
concluding remarks are summarized in Sect.~6.

\sect{High-energy QCD and Heisenberg spin magnet}

Let us consider elastic hadron-hadron scattering in QCD at high energy,
$s$, and fixed transferred momentum, $t$, and let us replace for simplicity
nonperturbative hadronic states by perturbative onium states built from two
heavy quarks with mass $M$. In this case, the scattering amplitude can be
calculated in perturbative QCD as a sum of Feynman diagrams describing the
multi-gluon exchanges between heavy quarks in the $t-$channel \ci{BFKL}.
In the Regge limit, $s\gg -t\,, M^2$, the perturbative expressions for
the Feynman diagrams involve large logarithmic corrections, $\as^n \ln^m s$,
which can be classified into the leading logs ($n=m$), next-to-leading
logs ($n=m+1$) etc. and which have to be resummed to all orders in $\as$.
We recall, however, that one of the basic concepts of perturbative approach
in QCD is that one considers quarks and gluons as elementary excitations and,
assuming that their interaction is small, takes it into account as a
perturbative expansion in $\as$. The necessity of resuming perturbative
series indicates that the description of a hadron-hadron scattering in
terms of ``bare'' quarks and gluons is not appropriate anymore in the Regge
limit and one has to identify instead a new collective degrees of freedom,
in terms of which the QCD dynamics will be simpler.

In the Regge limit, due to remarkable property of the Reggeization in QCD
\ci{BFKL}, reggeized gluons, or Reggeons, become a new collective excitations.
Although the Reggeon is built from an infinite number of ``bare'' gluons it
behaves as a point-like particle. It carries the color charge of a gluon
and it has a 4-momentum $k_\mu=(k_+,k_-,k_\perp)$ with the longitudinal
components $k_\pm$ belonging to the plane defined by the momenta of scattering
hadrons and $k_\perp$ being 2-dimensional transverse momentum.
The hadrons scatter each other by exchanging Reggeons in the
$t-$channel and the interaction between Reggeons is described by the
effective $S-$matrix, which should be obtained out of QCD in the Regge limit.

The Reggeon $S-$matrix has the following remarkable properties. Reggeons
propagate in the $t-$channel between two hadrons and due to strong ordering
of their longitudinal momenta the Reggeon rapidity $y=\ln\frac{k_+}{k_-}$
can be interpreted as a ``time'' in the $t-$channel \ci{Gr}. Then, the
Reggeon $S-$matrix describes the propagation of the Reggeons in the time
$y$. Interacting with each other Reggeons change their 2-dimensional
transverse momenta $k_\perp$ and color charge. Therefore, although 
Reggeons ``live'' in 4-dimensional Minkowski space-time,
their $t-$channel evolution is described by the effective $(2+1)-$dimensional
$S-$matrix \ci{Leff,VV}. The exact expression for the Reggeon $S-$matrix is
unknown yet and one may try to evaluate it in different approximations.
In what follows we will study the Reggeon $S-$matrix in the generalized LLA
\ci{BKP}. In this approximation one preserves the unitarity of the $S-$matrix
in the direct ($s-$, $t-$ and $u-$) channels but not in the subchannels.

\subsection{Generalized LLA}

The hadron-hadron (or onium-onium) scattering amplitude is given in the 
generalized LLA by the sum of effective Reggeon Feynman diagrams
\ci{BKP} in which, first,
the number of Reggeons propagating in the $t-$channel is preserved
(no creation/annihilation of Reggeons) and, second, only two Reggeons
could interact with each other at the same moment of ``time'' $y$.
Therefore, according to the number of Reggeons, $N$, the scattering
amplitude can be decomposed as
$$
A(s,t)=\sum_{N=2}^\infty \as^{N-2} A_N(s,t)\,,
$$
where the $N-$th term describes the evolution in the $t-$channel
of the system of $N$ Reggeons with a pair-wise interaction. The
amplitude $A_N(s,t)$ satisfies the Bartels-Kwiecinski-Praszalowicz
equation \ci{BKP}, which can be interpreted as a Bethe-Salpeter equation
for the $N$ Reggeon scattering amplitude. Its solutions define the
color-singlet $N$ Reggeon compound states, perturbative QCD
Pomerons and Odderons, whose contribution to the scattering amplitudes
takes the standard Regge form
$$
A_N(s,t)= is \sum_{\{q\}} \beta_A^{\{q\}}(t) \beta_B^{\{q\}}(t)\,
s^{E_{N,\{q\}}}\,,
$$
where indices $A$ and $B$ refer to the scattered hadrons.
Here, $E_{N,\{q\}}$ is the energy of the $N$ Reggeon compound state.
It is defined as an eigenvalue of the $N$ Reggeon hamiltonian
\be
{\cal H}_N\ket{\chi_{N,\{q\}}}= E_{N,\{q\}}\ket{\chi_{N,\{q\}}}\,,
\lab{BKP}
\ee
with $\{q\}$ being some set of quantum numbers parameterizing all
possible solutions. The residue functions $\beta_A^{\{q\}}$
and $\beta_B^{\{q\}}$ measure the overlapping between hadronic 
wave functions and the wave functions of the $N$ Reggeon compound
states
$$
\beta_A^{\{q\}}(t)= \vev{A | \chi_{N,\{q\}}}\,.
$$
The pair-wise Reggeon hamiltonian ${\cal H}_N$ acts in \re{BKP}
on 2-dimensional transverse momenta of $N$ Reggeons
and the relation \re{BKP} has the form of a (2+1)--dimensional Schr\"odinger
equation. Let us change the representation from 2--dimensional transverse
Reggeon momenta space $k_\perp$ to 2-dimensional impact parameter space
$b_\perp=(\xi,\zeta)$ and define the holomorphic and antiholomorphic complex
coordinates of Reggeons as
\be
z_j=\xi_j+i\zeta_j\,,\qquad \bar z_j=\xi_j-i\zeta_j\,,\qquad j=1,2,...,N\,.
\lab{imp}
\ee
Remarkable property of the Reggeon hamiltonian ${\cal H}_N$ is that being
expressed in terms of holomorphic and antiholomorphic Reggeon coordinates,
it splits in the multi-color limit, $\as N_c=\mbox{fixed}$ and
$N_c\to \infty$, into the sum of holomorphic and antiholomorphic
hamiltonians \ci{Lip2}
\footnote{For $N=2$ and $N=3$ this relation is exact even for finite $N_c$.}
$$
{\cal H}_N=\frac{\as N_c}{4\pi}\lr{H_N+\widebar H_N} + \CO(N_c^{-2})\,,
\qquad
[H_N,\widebar H_N]=0\,,
$$
where $H_N$ and $\widebar H_N$ act on holomorphic and antiholomorphic
Reggeon coordinates, respectively. Since the hamiltonians $H_N$ and
$\widebar H_N$ commute with each other, the original (2+1)--dimensional
Schr\"odinger equation \re{BKP} can be replaced by the system of
holomorphic and antiholomorphic (1+1)--dimensional Schr\"odinger equations
\be
H_N\ket{\varphi_{N,\{q\}}} = \varepsilon_{N,\{q\}} \ket{\varphi_{N,\{q\}}}
\,,\qquad
\widebar{H}_N\ket{\widebar\varphi_{N,\{q\}}} =
\widebar\varepsilon_{N,\{q\}} \ket{\widebar\varphi_{N,\{q\}}}\,,
\lab{hol}
\ee
which define the spectrum of the Reggeon compound states as follows
\be
E_{N,\{q\}}
=\frac{\as N_c}{4\pi}\lr{\varepsilon_{N,\{q\}}+\widebar\varepsilon_{N,\{q\}}}
\,,\qquad
\chi_{N,\{q\}}(z,\bar z)=
\varphi_{N,\{q\}}(z)\,
\widebar\varphi_{N,\{q\}}(\bar z)\,,
\lab{state}
\ee
where $z$ and $\bar z$ denote the full set of $N$ Reggeon holomorphic and
antiholomorphic coordinates.
Thus, in the generalized LLA perturbative (3+1)--dimensional multi-color
QCD is effectively replaced by the system of (1+1)--dimensional
Schr\"odinger equations \re{hol}. The properties of two 
equations in \re{hol} are similar and in what follows we will study
only the holomorphic Schr\"odinger equation in \re{hol} for fixed
number of Reggeons $N$.

\subsection{Integrability}

The holomorphic hamiltonian $H_N$ describes the nearest-neighbour
interaction between $N$ Regge\-ons with holomorphic coordinates $z_1$,
$...$, $z_N$ and periodic boundary conditions $z_{k+N}=z_k$
\be
H_N = \sum_{j=1}^{N+1} H_{j,j+1}
\lab{RH}
\ee
and $H_{N,N+1}=H_{N,1}$. Here, the interaction hamiltonian between two
Reggeons, $H_{j,j+1}$, is given by the BFKL kernel \ci{BFKL,Lip1}
\be
H_{12}\equiv H(z_1,z_2) = -\psi(-J_{12}) - \psi(1+J_{12}) +2\psi(1)\,,
\lab{ker}
\ee
where $\psi(x)=\frac{d\ln\Gamma(x)}{dx}$ and the operator $J_{12}$
is defined as an operator solution of the equation
$$
J_{12} (1 + J_{12}) = - (z_1-z_2)^2 \partial_1 \partial_2
$$
with $\partial_k=\partial/\partial z_k$. We stress that the operator
\re{ker} was originally derived \ci{BFKL} from the analysis of Feynman
diagrams contributing to the Regge asymptotics of the onium-onium scattering
amplitude in perturbative QCD. Therefore, it was quite unexpected to find
\ci{FK,Lip} that it coincides with the hamiltonian of 1-dimensional XXX
Heisenberg magnet of spin $s=0$ corresponding to the principal series
representation of the $SL(2,\IC)$ group. The number of sites of the spin
chain is equal to the number of Reggeons.

Among other things this equivalence means that the holomorphic Schr\"odinger
equation \re{hol} possesses the family of hidden mutually commuting
conserved charges \ci{FK,Lip}
\be
[q_k, q_j] = [q_k, H_N] = 0
\lab{com}
\ee
and their number is large enough for the system \re{hol} to be completely
integrable. To identify the conservation laws for the system of $N$
interacting Reggeons one applies the Quantum Inverse Scattering Method 
\ci{QISM} and uses well-known constructions for the Heisenberg magnet
\ci{XXX,R-m}. Namely, to each Reggeon we assign the (auxiliary) Lax operator
\be
L_k = \left(\begin{array}{cc} \lambda + i S_k^3 & iS_k^- \\
                               iS_k^+ & \lambda - i S_k^3
            \end{array}\right)
    = \lambda\cdot\II + i \left({1\atop z_k}\right)\otimes (z_k, -1)\partial_k
\,,
\lab{Lax}
\ee
where $\lambda$ is an arbitrary complex parameter and $S_k^\alpha$ are
spin $s=0$ generators of the $SL(2)$ group acting on the holomorphic
coordinates of the $k-$th Reggeon
$$
S_k^3 = z_k  \partial_k\,,\qquad
S_k^+ = z_k^2\partial_k\,,\qquad
S_k^- = -\partial_k\,.
$$
Then, we construct the monodromy matrix
\be
T(\lambda) = L_N(\lambda) L_{N-1}(\lambda) ... L_1(\lambda)
           = \left(\begin{array}{cc} A(\lambda) & B(\lambda) \\
                                      C(\lambda) & D(\lambda)
                     \end{array}\right)\,,
\lab{T}
\ee
where the operators $A$, $B$, $C$ and $D$ act on the holomorphic
coordinates of $N$ Reggeons and satisfy the Yang-Baxter equation
\ci{R-m,XXX}. Finally, we obtain the transfer matrix as
\be
\Lambda(\lambda) = \tr T(\lambda) = A(\lambda) + D(\lambda)
\lab{trT}
\ee
and verify, using \re{Lax} and \re{T}, that $\Lambda(\lambda)$ is a 
polynomial of the degree $N$ in $\lambda$
\be
\Lambda(\lambda) = 2\lambda^N + q_2 \lambda^{N-2} + ... + q_N\,,
\lab{Lam}
\ee
with the coefficients $q_k$ being operators acting on holomorphic
coordinates of the Reggeons. It immediately follows from the Yang-Baxter
equation for the monodromy matrix $T(\lambda)$ that the operators
$q_2$, $...$, $q_N$ satisfy the relations \re{com}
\footnote{Notice that in contrast with the spin$-\frac12$ Heisenberg
magnet and 1-dimensional Toda chain \ci{H}, the hamiltonian $H_N$ does
not enter into the expansion of the auxiliary transfer matrix \re{trT}.
To obtain $H_N$ one has to construct the fundamental transfer matrix.}.

Thus, we identify the operators $q_2$, $...$, $q_N$ as $N-1$ conserved
charges for the system of $N$ interacting Reggeons. To match the
number of conservation laws with the number of Reggeons, one needs one 
more conserved charge. It is easy to see that
the remaining $N-$th charge is associated with the center-of-mass motion
of the $N$ Reggeon compound state and it is equal to the total Reggeon
momentum
$$
P=\pi_1+\pi_2+...+\pi_N = i (S^-_1+S^-_2+...+S^-_N)
$$
with $\pi_j=-i\partial_j$ being the holomorphic component of
2-dimensional transverse momentum of the $j-$th Reggeon. The explicit 
expressions for the operators $q_k$ can be obtained from \re{Lax},
\re{T} and \re{Lam} as \ci{FK,Lip}
\be
q_k=\sum_{N\ge j_1>j_2>...>j_k\ge 1}
   i^k z_{j_1j_2}z_{j_2j_3} ... z_{j_kj_1}
   \partial_{j_1} \partial_{j_2} ... \partial_{j_k}
\lab{qk}
\ee
with $z_{jk}=z_j-z_k$. They can be interpreted as higher Casimir
operators of the $SL(2)$ group and their appearance can be traced
back to the invariance of the Reggeon hamiltonian \re{RH} under
the M\"obius transformations
\be
z_k\to \frac{az_k+b}{cz_k+d}\,,
\lab{Mob}
\ee
where $ac-bd=1$. In particular,
\be
q_2=\sum_{N\ge j >k \ge 1} z_{jk}^2\partial_j\partial_k=-h(h-1)
\lab{q2}
\ee
is the quadratic Casimir operator and its eigenvalue $h$ defines the
conformal weight of the holomorphic wave function $\varphi_N(z_1,...,z_N)$
of the $N$ Reggeon compound state \re{state}. The Reggeon wave function
belongs to the principal series representation of the $SL(2,\IC)$ group
and, as a result, its conformal weight is quantized
\be
h=\frac{1+m}2+i\nu,\,,\qquad m = \IZ\,,\quad \nu=\IR\,.
\lab{h-q}
\ee
Here, integer $m$ defines the Lorentz spin of the Reggeon state \re{state},
corresponding to the rotations in the 2-dimensional impact parameter space,
$\chi_N \to \e^{im\alpha}\chi_N$ as
$z_j \to \e^{i\alpha} z_j$ and $\bar z_j \to \e^{-i\alpha} \bar z_j$.
The quantization conditions for the remaining charges $q_3$, $...$, $q_N$
are much more involved \ci{Qua} and their interpretation in terms of
the properties of Reggeon states will be given in Sect.~5.

Once we identified the complete set of conservation laws,
the Reggeon hamiltonian becomes a rather complicated function of the
conserved charges
\footnote{Due to invariance of the Reggeon hamiltonian under M\"obius
transformations \re{Mob}, $H_N$ does not depend on the total momentum $P$.},
$H_N=H_N(q_2,...,q_N)$, and one can replace the original Schr\"odinger
equation \re{hol} by a simpler problem of simultaneous diagonalization
of the operators $P$, $q_2$, $...$, $q_N$. Their eigenvalues form a complete
set of quantum numbers parameterizing the $N$ Reggeon compound states,
$\varphi_{N,\{q\}}(z_1,...,z_N)$.

\sect{Separation of variables}

Diagonalization of the conserved charges can be performed using
the Separation of Variables (SoV) \ci{FM,Guz,SoV}. The operators $P$,
$q_2$, $...$, $q_N$ depend on the coordinates and momenta, $z_k$ and
$\pi_k=-i\partial_k$, respectively, of $N$ Reggeons and their
diagonalization is reduced to solving of a complicated system of
$N$ coupled partial differential equations for the holomorphic
wave function of the $N$ Reggeon state. Instead of dealing with
this system we perform a unitary transformation
\be
(z_j,\pi_j) \to (x_j=U^\dagger z_j U, p_j = U^\dagger \pi_j U )\,,
\qquad U=U(\{z_k,\pi_k\})
\lab{U}
\ee
in order to replace the original set of Reggeon coordinates and
momenta, $(z_j,\pi_j)$, by a new canonical set of
{\it separated variables\/}
\be
[x_j,x_k]=[p_j,p_k]=0\,,\qquad [p_j,x_k]=-i\delta_{jk}\,,
\lab{h=1}
\ee
in terms of which the following relations hold
\ba
\Phi(x_j,p_j;q_2,...,q_N)\ket{\varphi_{N,\{q\}}}&=&0\,,\quad j=1,2,...,N-1\,,
\lab{sep}
\\[3mm]
(p_N-P)\ket{\varphi_{N,\{q\}}}&=&0\,.
\nonumber
\ea
Here, the number of pairs $(x_j,p_j)$ is equal to the number of
Reggeons, $N$. One of the pairs, $(x_N,p_N)$, describes the
center-of-mass motion of the $N$ Reggeon state
\be
x_N=\frac1{N}{\sum_{j=1}^Nz_j}\,,\qquad
p_N=\sum_{j=1}^N \pi_j\,,
\lab{cm}
\ee
while the definition of the remaining separated coordinates and the
functions $\Phi$ will be given below. In relation \re{sep}, the
operators are ordered inside the function $\Phi$ as they are enlisted
and the conserved charges $q_k$ can be replaced by their eigenvalues 
corresponding to the $N$ Reggeon compound state $\varphi_{N,\{q\}}$.

The remarkable property of the separated coordinates is that in the
$x-$representation, $p_j=-i\partial/\partial x_j$, the relations \re{sep}
define the system of $N$ separated differential equations for the
wave function of the $N$ Reggeon state
$\varphi_{N,\{q\}}=\varphi_{N,\{q\}}(x_1,...,x_N)$
and its solution takes the following factorized form
\be
\varphi_{N,\{q\}}(x_1,...,x_N)= Q(x_1) Q(x_2) ... Q(x_{N-1}) \exp(iPx_N)\,,
\lab{wf}
\ee
where $P$ is the total momentum of the state and a function $Q(x)$
satisfies the Baxter equation
\be
\Phi(x,-i\partial_x;q_2,...,q_N) Q(x)=0\,.
\lab{Q}
\ee
To obtain the wave function in terms of the holomorphic Reggeon coordinates
one has to go back from $x-$ to $z-$representation in \re{U} by applying
the unitary transformation $U$. The resulting expression can be found in
\ci{FK}.

\subsection{Baxter equation}

Let us construct the separated variables for the system of $N$ Reggeons.
Before doing this we would like to notice that the SoV method has been
extensively developed in application to 1-dimensional {\it classical\/}
integrable systems. In this case, as it was shown in many examples
\ci{NMPZ}, poles of a properly normalized Baker-Akhiezer function
provide the set of separated coordinates. Having expressions for the
separated variables in the classical system, one may try to generalize 
them to the corresponding quantum integrable model. However, one of the 
peculiar features of the holomorphic Schr\"odinger equation for the $N$ 
Reggeon compound states is that it describes 1-dimensional essentially 
quantum integrable system and one has to determine the separated
variables without knowing their classical analogs.

Constructing the separated variables for the $N$ Reggeon state we follow
the approach developed by Sklyanin \ci{SoV}. Namely, we define the separated
coordinates $x_1$, $...$, $x_{N-1}$ as operator zeros of the operator
$B(\lambda)$ entering the expression \re{T} for the monodromy matrix.
According to \re{Lax} and \re{T}, $B(\lambda)$ is a polynomial
of the degree $N-1$ in $\lambda$ and it can be represented as
$$
B(\lambda)= p_N (\lambda-x_1) (\lambda-x_2) ... (\lambda-x_{N-1})\,,
$$
where $p_N$ was defined in \re{cm} and the ordering of
operators is unessential due to \re{h=1}. The definition of the momenta
$p_j$ conjugated to the coordinates $x_j$ takes a few steps.
Let us substitute $\lambda=x_j$ into \re{T}. Since the operators $A(\lambda)$,
$C(\lambda)$ and $D(\lambda)$ do not commute with $x_j$ we use the
prescription of ``substitution from the left'', $\lambda\mapsto x_j$.
This means \ci{SoV}, that for an arbitrary operator
$W(\lambda)=\sum_k \lambda^k W_k$
one first pulls out all powers of $\lambda$ to the left and then replaces
$\lambda$ by the operator $x_j$ as
$W\lr{\lambda\mapsto x_j}=\sum_k x_j^k W_k$. For
$\lambda\mapsto x_j$ the monodromy matrix \re{T} takes a triangle form 
with the diagonal elements given by
\be
A\lr{\lambda\mapsto x_j} = x_j^N \omega_j^+\,,\qquad
D\lr{\lambda\mapsto x_j} = x_j^N \omega_j^-\,,
\lab{omega}
\ee
where notation was introduced for the operators $\omega_j^+$ and
$\omega_j^-$.

Let us forget for a moment that we are dealing with operators and
replace all operators by the corresponding classical functions.
Then, one could identify the diagonal elements $x_j^N \omega_j^\pm$
of the monodromy matrix $T(\lambda)$ as its eigenvalues and write down
the characteristic equation for $\omega^\pm$ as
$$
\det(T(x)-x^N\omega)=x^{2N}\omega^2-x^N \omega\ \tr T(x) +
\det T(x)=0\,.
$$
Using the definition of the transfer matrix, \re{trT},
and that classically
$\det T(\lambda)=\prod_{k=1}^N \det L_k(\lambda)$ $=\lambda^{2N}$
we obtain the relation
\be
\omega+\frac1{\omega} = x^{-N} \Lambda(x)
= 2 + \frac{q_2}{x^2}+ ... +\frac{q_N}{x^N}\,,
\lab{sc}
\ee
which can be interpreted as the level surface of the commuting integrals
of motion $q_k$, or spectral curve of the monodromy operator for some
classical Liouville integrable system \ci{NMPZ}. As we will show in the next
section, this system arises as a quasiclassical limit of the $N$ Reggeon
state.

Let us take now into account noncommutativity of operators and
establish the quantum analog of the spectral curve \re{sc}. Using
the Yang-Baxter equation for the monodromy matrix \re{T} one can verify
the following relations for the operators $\omega^\pm_k$ \ci{SoV}
$$
[\omega_j^\pm,\omega_k^\pm]=[\omega_j^\pm,\omega_k^\mp]=0\,,
\qquad
\omega_j^+\omega_j^-=1\,,
\qquad
\omega_j^\pm x_k = (x_k\pm i \delta_{jk}) \omega_j^\pm\,.
$$
These relations suggest that the operators $\omega^\pm_j$ act on the
Reggeon wave function in the $x-$repre\-sen\-tation as shift operators and
they allow us to define the momentum $p_j$ conjugated to
$x_j$ in the separated variables as follows
\be
\omega_j\equiv \omega_j^+=\frac1{\omega_j^-}= \pm \exp(-p_j)\,,
\qquad
\omega_k Q(x_j) = \pm Q(x_j+i\delta_{jk})\,.
\lab{p}
\ee
We notice that the canonical commutation relations \re{h=1} do not allow
to fix a sign ambiguity in the definition of the momentum.

To establish the relations \re{sep} in the separated variables $(x_j,p_j)$,
we substitute $\lambda\mapsto x_j$  into the definition \re{trT}
of the transfer matrix and use \re{omega} to get
\be
\omega_j+\frac1{\omega_j}
=x_j^{-N}\Lambda(x_j)=2+ x_j^{-2} q_2 + ... + x_j^{-N}q_N
\,,
\qquad
\lab{sc-q}
\ee
where $j=1$, $...$, $N-1$ and
the ordering of the operators $x_j$ and $q_k$ is important since
they do not commute with each other. The operator identity \re{sc-q} can be
considered as a quantum analog of the spectral curve \re{sc}. It relates
a pair of the separated variables $(x_j,p_j)$ to the set of the conserved
charges and being applied to the wave function of the $N$ Reggeon compound
state it leads to \re{sep}, provided that
\be
\Phi(x_j,p_j;q_2,...,q_N)= x_j^{-N} \Lambda(x_j) - 2\cosh p_j\,.
\lab{Phi}
\ee
Here, we used \re{p} with a plus sign and took into account
the possibility to change a sign by introducing an ambiguity
$p_j\to p_j+i\pi$ in the definition of the momentum%
\footnote{Similar to the situation in the Toda chain \ci{FM},
the sign ambiguity can be fixed by requiring the momentum $p_j$ to
have real eigenvalues.}.
Substituting \re{Phi} into \re{Q} we find that the wave function $Q(x_j)$
satisfies the finite-difference equation
\be
x_j^{-N} \Lambda(x_j)\, Q(x_j) = Q(x_j+i)+Q(x_j-i)
\,, \qquad j=1,2,..., N-1\,,
\lab{Bax}
\ee
which has the same form as the equation for eigenvalues of the Baxter
$Q-$operator \ci{Q}. In this equation, $x_j$ denotes the eigenvalue of the
corresponding coordinate operator. Notice that the operators $q_k$
entering into the definition of the transfer matrix \re{Lam} have been
replaced in \re{Bax} by their eigenvalues corresponding to the $N$ Reggeon
compound state. As for any quantum system, the possible values of
$q_k$ are constrained by the quantization conditions which will
be discussed below.

Having solved the Baxter equation \re{Bax}, one can obtain the wave
function of the $N$ Reggeon compound state \re{wf}. Moreover, the same
function $Q(x)$ controls the dependence of the holomorphic Reggeon
hamiltonian on the conserved charges and for a given set of quantum numbers
$q_2$, $...$, $q_N$ the holomorphic energy of the $N$ Reggeon state,
defined in \re{hol}, can be evaluated as \ci{FK,Bet}
\be
\varepsilon_N(q_2,...,q_N)
= i\frac{d}{dx}\ln\frac{Q(x-i)}{Q(x+i)}\bigg|_{x=0}\,.
\lab{energy}
\ee

The Baxter equation \re{Bax} alone does not allow us to find its solution
$Q(x)$. One has to specify additionally the spectrum of the
operators $x_1$, $...$, $x_N$ and provide the appropriate boundary
conditions for the function $Q(x)$. It is clear that the possible
eigenvalues of the coordinate operators $x_k$ depend on the class
of functions $Q(x)$ on which they act. For example, the spectrum of
$x_j$ is discrete for the spin$-\frac12$ Heisenberg magnet, while for
the Toda chain the eigenvalues of $x_j$ take continuous real values
\ci{FM,Guz,SoV}. To find the spectrum of the operators $x_j$ for the
Reggeon states we have to impose additional conditions on the functions $Q(x)$.
To this end we notice that the solution of the Baxter equation $Q(x)$
defines the wave function of the Reggeon compound state \re{wf} and the
above conditions should follow from the requirement for its norm to be
finite. Although the explicit expression for the norm is known \ci{FK,Lip},
the general form of the resulting constraints on the functions $Q(x)$
was not found yet, except of the subclass of the so-called
polynomial solutions of the Baxter equation \ci{Qua}, corresponding
to the special values of quantized conformal weight \re{h-q}.

\subsection{Polynomial solutions of the Baxter equation}

The subclass of polynomial solutions covers all normalizable solutions
of the Baxter equation corresponding to the {\it integer\/} positive
conformal weight of the $N$ Reggeon compound states
\be
h = \IZ_+\,,\qquad
h \ge N\,,
\lab{h+}
\ee
where one put $\nu=0$ and $m \ge 2N-1$ to be odd in \re{h-q}.
In this case, as was shown in \ci{Qua}, the normalizable solutions $Q(x)$
of the Baxter equation become polynomials of the degree $h$ in $x$
with $N$ time degenerate root $x=0$
\be
Q(x)=x^N \prod_{j=1}^{h-N} (x-\lambda_j)\,.
\lab{poly}
\ee
Each solution is characterized by the set of nonvanishing roots
$\{\lambda_j\}$. Having the explicit expressions for the roots, one
can easily evaluate from \re{Bax} and \re{energy}
the quantum numbers $q_k$ and the corresponding
energy of the $N$ Reggeon states. It follows from the Baxter
equation \re{Bax} that $\lambda_j$ satisfy the Bethe equations for the
XXX magnet of spin $s=0$ \ci{Bet}. Their study reveals the following
interesting properties of the polynomial solutions of the Baxter
equation \ci{Bet,Qua}.

For any given integer $h$ in \re{h+}, the space of polynomial
solutions is finite-dimensional. The possible values of 
roots $\lambda_k$, as well as the values of quantized $q_k$
and the energy $\varepsilon_N$, turn out to be real and simple
\be
\Im \lambda_j = \Im q_k = \Im \varepsilon_N = 0
\lab{real}
\ee
and they can be parameterized by integers $n_1$, $...$, $n_{N-2}$
\be
q_k=q_k(h;n_1,...,n_{N-2})\,,\qquad
\varepsilon_N=\varepsilon_N(h;n_1,...,n_{N-2})
\lab{int}
\ee
such that
$$
n_1,..., n_{N-2} \ge 0\,,\qquad \sum_{k=1}^{N-2} n_k \le h-N\,.
$$
The eigenvalues of the coordinate operators $x_1$, $...$, $x_{N-1}$ are
also real for the polynomial solutions and they occupy $N-1$ compact
nonoverlapping intervals on the real axis
\be
\Im x_j = 0\,,\qquad x_j \in [\sigma_{2j-1},\sigma_{2j}]\,.
\lab{inter}
\ee
The positions of the intervals, $\sigma_k$, depend on the values of quantized
$q_3$, $...$, $q_N$ and they satisfy the following relation
\be
\Lambda^2(\sigma_k)=4\sigma_k^{2N}\,.
\lab{sigma}
\ee
The possible values of roots $\lambda_k$ belong to the same intervals
\re{inter} and their distribution on the real axis was found
in the limit of large integer conformal weight, $h \gg 1$.
Moreover, in the large $h$ limit the following scaling holds
\be
\lambda_j = \CO(h)\,,\qquad q_2 = \CO(h^2)\,, \ ...\ , \
q_N = \CO(h^N)\,,\ \ \
\varepsilon_N = \CO(\ln h)\,,
\lab{sca}
\ee
which allows to develop the asymptotic expansion of  $q_k$ and
$\varepsilon_N$ in inverse powers of the conformal weight
\be
q_k = h^k \sum_{l=0}^\infty q_k^{(l)}(n) \, h^{-l}\,,\qquad
\varepsilon_N =-2N\ln h + \sum_{l=0}^\infty
\varepsilon_N^{(l)}(n) \, h^{-l}\,.
\lab{as-exp}
\ee
where $k=3$, $...$, $N$ and the coefficients $q_k^{(l)}$ and
$\varepsilon_N^{(l)}$ depend on the integers $n_k$. For $N=2$ Reggeon
states all coefficients are
known exactly and for $N=3$ Reggeon states they were calculated up to
$\CO(h^{-8})$ order \ci{Qua}.

Comparing the asymptotic expansions \re{as-exp} for $N=2$ and $N\ge 3$
Reggeon states one can discover two important differences \ci{Qua}. First,
the coefficients $\varepsilon_N^{(l)}$ grow as factorials to higher
orders in $1/h$. The asymptotic series for the energy of $N=2$ Reggeon states
is Borel summable, but for $N\ge 3$ the properties of the same series
are drastically changed and it becomes non Borel summable.
Second, for $N\ge 3$ Reggeon states one has to study additionally the
properties of the ``higher'' quantum numbers $q_3$, $...$, $q_N$. One
finds that for $N\ge 3$ Baxter equation among all possible polynomial
solutions there are a few singular solutions, corresponding to the
situation, when either one of the intervals \re{inter} is shrinking into a
point, or two intervals merge, $\sigma_{j-1}=\sigma_{j}$. On the quantum
moduli space of $q_3$, $...$, $q_N$ the singular solutions are described
by the ``critical'' $(N-3)-$dimensional hypersurface
\be
\Sigma_N(q_3^{\rm crit},...,q_N^{\rm crit})=0\,.
\lab{crit}
\ee
For example, for $N=3$ Reggeon states the critical values of $q_3$ are
just three points given by
\be
q_3^{\rm crit}=0 \quad \mbox{and} \quad
\pm \frac{h^3}{\sqrt{27}}\lr{1+\CO(h^{-1})}\,.
\lab{q3-cr}
\ee
As we will show in Sect.~4, both properties are closely related to
each other and they can be simply understood using the properties
of hyperelliptic Riemann surfaces determined by the complex curve
\re{sc}.

The polynomial solutions of the Baxter equation are closely related to the
different systems of orthogonal polynomials \ci{Bet}. For $N=2$ the
exact solution of the Baxter equation was identified as a continuous
Hahn symmetric polynomial. For $N\ge 3$ more complicated systems of
orthogonal polynomials arise and their properties can be studied using
the approach developed in \ci{SZ}.

\subsection{Quasiclassical limit}

The derivation of the Baxter equation in Sect.~3.1 was based on the
close relation between Heisenberg spin magnet and the system of $N$
interacting Reggeons in multi--color QCD. It is not surprising therefore
that similar equations appear in different 1-dimensional quantum
integrable systems. Namely, applying the SoV to the periodic Toda chain with
$N$ sites one obtains the Baxter equation in the form \ci{Guz,SoV,PG}
\be
\Lambda_{\rm Toda}(x_j)\, Q(x_j)=Q(x_j+i\hbar)+Q(x_j-i\hbar)
\,,\qquad j=1,2,...,N-1\,,
\lab{Toda}
\ee
where $\Lambda_{\rm Toda}$ is the corresponding transfer matrix and
$\hbar$ is the Planck constant. Comparing \re{Bax} and \re{Toda}
we notice a trivial but important difference between two models --
the Baxter equation for the $N$ Reggeon state, as well as the canonical
commutation relations \re{h=1}, do not involve the Planck constant.
Indeed, the Planck constant enters into the definition of the
XXX Heisenberg magnet as a coefficient in front of the
second term in the Lax operator \re{Lax}. The Reggeon Lax operator \re{Lax}
and, as a consequence, the Baxter equation \re{Bax} correspond to the
special case $\hbar =1$.

Thus, the system of $N$ Reggeons does not have any natural small parameter,
which would allow us to perform the quasiclassical limit,
$Q(x)=\exp(\frac{i}{\hbar}S(x))$ as $\hbar\to 0$, and obtain
its classical integrable analog. Nevertheless, there is a simple way
of introducing a small parameter $\eta$ into the Baxter equation
similar to the one used in asymptotic solutions of the Painlev\'e type
I equation \ci{P-I}. Let us rescale in \re{Bax} the collective coordinate as
$x\to x/\eta$ and introduce notations
\be
f(x)=\eta^h Q(x/\eta)\,,
\qquad
\hat q_k = q_k \eta^k\,.
\lab{hat}
\ee
Then, one could identically rewrite the Baxter equation \re{Bax} as
\be
f(x+i\eta)+f(x-i\eta)=x^{-N} \Lambda(x,\{\hat q_k\}) f(x)\,,
\lab{B-f}
\ee
where the transfer matrix is given by \re{Lam} with $q_k$ replaced by
$\hat q_k$ defined in \re{hat}. Following \ci{P-I}, let us look for
the solutions of the Baxter equation in the form
\be
f(x)=\exp\lr{\frac{i}{\eta}S(x,\{\hat q_k\})}\,,\qquad
S(x,\{\hat q_k\})=S_0(x)+\eta S_1(x) + ...  \ ,
\lab{exp}
\ee
where each term is assumed to be uniformly bounded and the expansion
to be convergent.

The asymptotic expansion \re{exp} of the solutions of the Baxter equation
naturally appears in the limit $\eta\to 0$ and $\hat q_k=\mbox{fixed}$.
According to \re{hat} and \re{q2}, this limit corresponds to the large values
of the conformal weight, $h\sim 1/\eta \to \infty$, and it was studied
in \ci{Qua}. Using \re{poly}, \re{hat} and \re{exp} we can express $S$ in
terms of the roots $\lambda_j$ as
\be
S(x;\{\hat q_k\})=-i\eta\sum_{j=1}^h \ln(x-\lambda_j\eta)
\,,
\lab{S}
\ee
where $\lambda_{h-N+1}=...=\lambda_h=0$. Then, for $h\sim 1/\eta \to
\infty$ the property of roots \re{sca} implies the scaling
$\lambda_j\eta=\CO(\eta^0)$ and
$S(x;\{\hat q_k\})=\CO(\eta^0)$, which leads to the asymptotic
expansion \re{exp}.

Substituting the ansatz \re{exp} into \re{B-f} and expanding the both sides
of the Baxter equation \re{B-f} to order $\eta$ one obtains \ci{PG}
\be
2\cosh S_0'(x)=x^{-N}\Lambda(x;\{\hat q_k\})\,,\qquad
S_1'(x)=\frac{i}2 S_0''(x)\coth S_0'(x)\,,
\lab{S0}
\ee
where prime denotes a derivative with respect to $x$. Then, the solution
of the Baxter equation \re{B-f} is given by
\be
f(x)=\frac1{\left[\sinh S_0'(x)\right]^{1/2}}
\exp\lr{\frac{i}{\eta}\int^x d S_0(x)+ \CO(\eta)}\,.
\lab{f}
\ee
Comparing \re{B-f} with the Baxter equation for the Toda chain \re{Toda}
we observe that the parameter $\eta$ plays a role of the
Planck constant. The ansatz \re{f} takes the form of the
WKB expansion with the leading term $S_0(x)$ being the ``Reggeon
classical action''. Let us study the classical mechanics of the $N$
Reggeon system governed by the action $S_0(x)$.

\sect{Quasiclassical Reggeon states as KP/Toda solitons}

Let us introduce notation for the function
\be
\omega(x)=\exp(S_0'(x))
\lab{S0-1}
\ee
and rewrite the first equation in \re{S0} as
\be
\omega+\frac1{\omega}=x^{-N} \Lambda(x;\{\hat q_k\})\,.
\lab{spe}
\ee
This expression is the quasiclassical analog of the operator relation
\re{sc-q}. We notice that the variable $x$ appears in \re{B-f} and
\re{spe} as a collective coordinate of the Reggeons and for polynomial
solutions it takes values inside one of the $N-1$ compact intervals
on the real axis \re{inter}. The classical analog of the Reggeon momentum
in the separated coordinates, $[x,p]=i\eta$, can be obtained from
\re{p} as
\footnote{The same quantity can be interpreted \ci{G} as a quasimomentum
for the auxiliary first-order difference linear problem corresponding to
the Reggeon Lax operator, $L_n \psi_n(E) = E \psi_{n+1}(E)$, with the
periodicity condition $\psi_{N+n}=\e^{p(E)}\psi_n$.}
\be
p=\ln |\omega(x)|\,,
\lab{p-cl}
\ee
where $\omega$ takes positive and negative values in \re{spe}
and the modulus is needed for momentum $p$ to be real. Using \re{spe}
and taking into account that $\omega$ is real on the classical trajectories,
we obtain the relations which define the intervals of the classical
motion of Reggeons as
\be
\Lambda^2(x_j;\{\hat q_k\}) \ge 4 x_j^{2N} \,,\qquad
j=1,2,...,N-1\,,
\lab{band}
\ee
where index refers to the $j-$th allowed band. This relation is in
agreement with \re{sigma}.

\subsection{Spectral curve}

Let us continue the relation \re{spe} into the complex domain. For
$\eta=1$ it coincides with the spectral curve \re{sc} and
replacing
\be
y=x^N\lr{\omega-\frac1{\omega}}
\lab{y}
\ee
one can rewrite it in the conventional form as
\be
\Gamma_N: \qquad y^2=\Lambda^2(x)-4 x^{2N} = 4\hat q_2 P_{2N-2}(x)
\lab{GamN}
\ee
where $P_{2N-2}(x)=x^{2N-2}+...\ $ is a polynomial with the coefficients
depending on the quantum numbers $\hat q_k$ of the $N$ Reggeon
state. The curve \re{GamN} determines the hyperelliptic Riemann
surface $\Gamma_N$  equipped with a meromorphic differential $dS_0$
defined in \re{S0}. The genus of the Riemann surface, $g=N-2$, depends
on the number of Reggeons inside the compound state. For $N=2$ Reggeon state,
the well-known BFKL Pomeron \ci{BFKL}, $\Gamma_2$ is a sphere,
\be
\Gamma_2: \qquad
y^2 = \hat q_2 (4 x^2+\hat q_2)
\lab{Gamma2}
\ee
while for $N=3$ Reggeon state, the QCD Odderon \ci{odd}, $\Gamma_3$
is a torus,
\be
\Gamma_3: \qquad
y^2= (\hat q_2 x+\hat q_3)(4 x^3 + \hat q_2 x + \hat q_3)\,.
\lab{Gamma3}
\ee
The branch points of the complex curve, $y=0$, correspond to $\omega=\pm 1$
(or $p=0$) in \re{y}. They coincide with the turning points of the
classical Reggeon trajectories and define the
boundaries of the allowed bands, $x=\sigma_j$, in \re{sigma}.
Then, the Riemann surface $\Gamma_N$ admits a representation in the
form of two sheets of the complex $x-$plane glued together along the
cuts running between the turning points $\sigma_{2j-1}$ and $\sigma_{2j}$
(see Fig.~1). One of the sheets will be called the upper and the other one
the lower sheet. Each point on the Riemann surface $\Gamma_N$ can be
parameterized as $Q=(x,\pm)$, where different signs correspond to the
upper and lower sheets.

\begin{figure}[ht]
\centerline{\epsffile{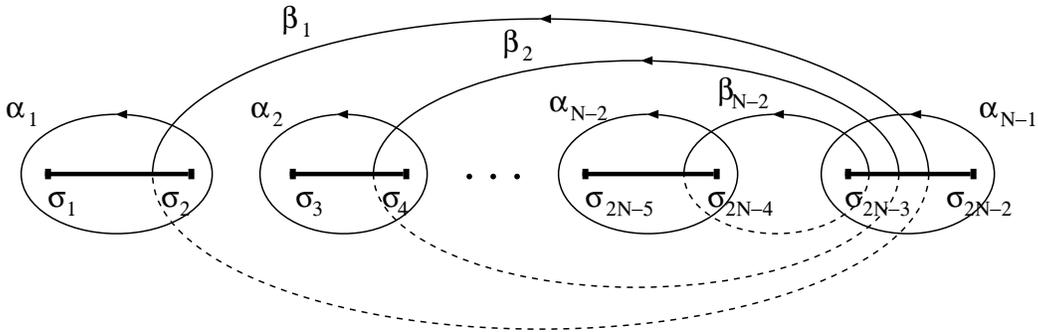}}
\caption{The definition of the canonical basis of cycles on the Riemann
surface $\Gamma_N$. The dotted line denotes the part of the $\beta-$cycles
belonging to the lower sheet. The $\alpha-$cycles correspond to the classical
motion of Reggeons over the allowed bands.}
\end{figure}

In the center-of-mass frame,
the classical trajectories of the Reggeons correspond to the
cycles over $N-1$ allowed bands on the Riemann surface%
\footnote{For polynomial solutions of the Baxter equation the
number of allowed bands should match the number of degrees of
freedom minus the motion of center-of-mass with the coordinate $x_N$.},
which we denote as $\alpha_k$ with $k=1$, $...$, $N-1$. By the definition,
for each point $Q$ on the cycle $\alpha_k$ the corresponding
Reggeon coordinate and momentum, $x_j$ and $p_j$, take real values.
It is easy to see from Fig.~1 that the sum of all $\alpha-$cycles is
homologous to zero
\be
\alpha_1+\alpha_2+...+\alpha_{N-1} \approx 0
\lab{0}
\ee
and one can choose the first $g=N-2$ $\alpha-$cycles to construct the
canonical basis of cycles on $\Gamma_N$: $\alpha_1$, $...$, $\alpha_{N-2}$,
$\beta_1$, $...$, $\beta_{N-2}$ with the intersection matrix
$$
\alpha_j \circ\alpha_k=\beta_j\circ\beta_k=0\,,\qquad
\alpha_j\circ \beta_k=\delta_{jk}\,.
$$
The definition of the $\beta-$cycles is shown in Fig.~1. Obviously,
the introduction of the $\beta-$cycles has a meaning only for $N\ge 3$
Reggeon states.

At this point we notice the important difference between $N=2$ and $N\ge 3$
Reggeon states -- the appearance of the moduli for the complex curve
$\Gamma_N$ for $g=N-2 \ge 1$. To identify the moduli we observe that
the curve \re{GamN} is invariant under transformation
\be
x\to \lambda x\,,\qquad\hat q_k \to \lambda^k \hat q_k\,,\qquad
y\to \lambda^N y\,,
\lab{sym}
\ee
which is induced by rescaling of the parameter $\eta$ in \re{hat}.
Let us introduce the moduli of the hyperelliptic curve $\Gamma_N$ for
$N\ge 3$ as
\be
u_{k-2}=\hat q_k/(-\hat q_2)^{k/2}=q_k/(-q_2)^{k/2}\,,\qquad
k=3,...,N\,,
\lab{hk}
\ee
where we used \re{hat} to eliminate the dependence on arbitrary parameter
$\eta$. Thus defined moduli are invariant under \re{sym} and
depend on the quantum numbers of the $N$ Reggeon states.
It is also convenient to introduce the following variable
$$
T=\lr{-\hat q_2}^{1/2} = \eta \lr{-q_2}^{1/2}\,.
$$
According to \re{real}, for polynomial solutions of the Baxter equation,
$u_1, ..., u_{N-2}$ and $T$ take {\it real\/} quantized values, such that
\be
\hat q_2=-T^2\,,\qquad
\hat q_k= T^k u_{k-2}\,,
\lab{T1}
\ee
with $k=3$, $...$, $N$.

The relation \re{S0-1} defines the following meromorphic 1--differential
on the hyperelliptic curve \re{GamN}
\be
dS_0 = dx \ln \omega(x) \cong -x \frac{d\omega}{\omega}
=\frac{N\Lambda(x)-x\Lambda'(x)}{y} dx\,.
\lab{dS0}
\ee
At the vicinity of two infinities on the upper and lower sheets of the
Riemann surface, $P_\pm$, the differential behaves as
$$
dS_0 \stackrel{x\to P_\pm}{\sim} \mp iT \frac{dx}{x}
$$
and we identify $dS_0$ as a dipole (unnormalized) differential of
the third kind \ci{D,NMPZ} with the residue
$\res_{P_\pm} dS_0=\frac1{2\pi i}\oint_{C(P_\pm)} dS_0=\pm iT$
at the first-order poles at $P_+$ and $P_-$ on the curve \re{GamN}.
Let us introduce by now standard notation for the periods of the
differential $dS_0$
\be
a_k=\oint_{\alpha_k} dS_0\,,\qquad
a^D_j=\oint_{\beta_j} dS_0\,.
\lab{per}
\ee
One can verify from \re{dS0}, \re{GamN} and \re{band} that $a_k$ and
$a^D_k$ take correspondingly real and pure imaginary values for the 
polynomial solutions. Using \re{0} one obtains that the sum of all 
$\alpha-$periods is given by the residue of $dS_0$ at infinity
\be
a_1+a_2+...+a_{N-1}=-2\pi i\res_{P_+}dS_0=2\pi T \,,
\lab{sum}
\ee
and only $N-2$ periods are linear independent. The periods are
functions of $T$ and $N-2$ moduli $u_k$ of the curve \re{GamN}
\be
a_k=a_k(T; u_1,...,u_{N-2})\,,\qquad k=1,2,..., N-2
\lab{ak}
\ee
and $a^D_k$ has a similar dependence.

Substituting the explicit expression \re{Lam} for the transfer matrix
into \re{dS0} we can expand the meromorphic differential $dS_0$ over
the basis of differentials of the first and the third kind on the curve 
$\Gamma_N$, $\omega_k$ and $d\Omega$, respectively, as
$$
dS_0=T\left[-2d\Omega+\sum_{k=1}^{N-2} (k+2) u_k d\omega_k\right]\,,
$$
where
\be
d\Omega = T \frac{dx\, x^{N-2}}{y}\,,\qquad
d\omega_k = T^{1+k} \frac{dx\, x^{N-2-k}}{y}\,,\qquad
(k=1, ..., N-2)\,.
\lab{diff}
\ee
Here, additional powers of $T$ were included to ensure invariance of the
differentials under transformation \re{sym},
which acts as $T\to \lambda T$ and leaves $u_k$ invariant.
The differentials \re{diff} depend only on the moduli $u_k$ but not on $T$.
Taking their linear combinations one can construct the canonical basis of
the holomorphic 1--differentials $d\hat\omega_k$ with $k=1$, $...$, $N-2$
\be
d\hat\omega_k= \sum_{j=1}^{N-2} U_{kj}(u) d\omega_j\,,\qquad
\oint_{\alpha_j} d\hat\omega_k=2\pi\delta_{jk}
\lab{bas-1}
\ee
and the normalized differential of the third kind, $d\hat\Omega$,
with the residue $\res_{P_\pm} d\hat\Omega=\pm 1$
at the first-order poles $P_\pm$
\be
d\hat\Omega= 2i d\Omega + i \sum_{j=1}^{N-2} V_j(u) d\omega_j
\,,\qquad
\oint_{\alpha_j} d\hat\Omega=0\,.
\lab{bas-3}
\ee
Here, $U_{kj}$ and $V_j$ are real coefficients depending on the moduli of
the curve $\Gamma_N$. Then, for given values of the periods \re{per} the
differential $dS_0$ can be expanded over the canonical basis as
\be
dS_0=iT d\hat\Omega + \sum_{k=1}^{N-2} \frac{a_k}
{2\pi}d\hat\omega_k\,.
\lab{can}
\ee
It is interesting to note that similar differential appears in the
Seiberg-Witten description of the low-energy effective action
of $N=2$ SUSY Yang-Mills theory \ci{IM}. The differential \re{can} defines
the asymptotic solution \re{f} of the Baxter equation, which being
substituted into \re{hat}, \re{wf} and \re{energy} determines the
spectrum of the $N$ Reggeon compound states.

\subsection{Hamiltonian flows}

The phase space for the system of $N$ Reggeons is given by the direct product
of the cycles $\alpha_j$ $(j=1,2,...,N-1)$ on the Riemann surface $\Gamma_N$
times the center-of-mass motion. The set of points $Q_1$, $...$,
$Q_{N-1}$ on the Riemann surface situated one each on the $\alpha-$cycles
corresponds to the real values of the canonical Reggeon coordinates
$(x_j,p_j)$ and provides the coordinates on the level surface
$\hat q_k=\mbox{const.}$ Let us consider the Hamiltonian
flow of the Reggeons on the curve $\Gamma_N$, generated by the hamiltonians
$\hat q_k$ with the canonical Poisson bracket in the separated variables
$(x_j,p_j)$ as
\be
\frac{\partial x_j}{\partial \tau_k}=\{x_j,\hat q_k\}
=\frac{\partial \hat q_k}{\partial p_j}\,,
\lab{Poi}
\ee
with $\tau_k$ being the corresponding ``times''. To calculate the
r.h.s.\ we use the properties \re{spe} and \re{p-cl} of the separated
variables to get
\be
2(\cosh p_j-1) x_j^N =\hat q_2 x_j^{N-2}+\hat q_3 x_j^{N-3}+...+\hat q_N
\lab{dis}
\ee
with $j=1$, $...$, $N-1$ and invert these relations to find the expressions
for the hamiltonians $\hat q_k$. It is convenient to introduce
$(N-1)\times(N-1)$ matrix $W$ inverse to the Vandermonde matrix
\be
(W^{-1})_{jk}=x_j^{N-k}\,,\qquad
W_{kj}=\frac{(-)^k}{m'(x_j)}\frac{\partial}{\partial x_j}t_{k-1}(x)
\,,
\lab{VdM}
\ee
where $1 \le j \le N-1$, $2 \le k \le N$, $m(x)=\prod_{j=1}^{N-1}(x-x_j)$ and
$t_k=\sum_{1\le j_1<...<j_k\le N-1}x_{j_1} ... x_{j_k}$
are symmetric polynomials.
Then, one obtains from \re{dis}
the expressions for the hamiltonians
\be
\hat q_k=2\sum_{j=1}^{N-1} W_{kj}(x) x_j^N (\cosh p_j-1)\,,
\lab{q-H}
\ee
which have a nonpolynomial kinetical part and which are very similar to
analogous Hamiltonians for the classical Toda chain of $N$ interacting
particles \ci{Guz,vM}. The hamiltonians \re{q-H}
form a commutative family with respect to the Poisson bracket \re{Poi}.
Substitution of \re{q-H} into \re{Poi} yields the equations of motion
for the Reggeon coordinates
\be
\frac{\partial x_j}{\partial \tau_k}
=2W_{kj}(x) x_j^N \sinh p_j
=W_{kj}(x) \sqrt{\Lambda^2(x_j)-4x_j^{2N}}\,,
\lab{em}
\ee
where \re{y} and \re{p-cl} were used. These relations are well-known
in the theory of finite-gap solutions of the KP/Toda systems as equations for
zeros (or poles) of the Baker-Akhiezer function \ci{NMPZ}.

The integration of the evolution equations \re{em} can be easily
performed by the Abel map \ci{NMPZ,D,Kr}.
Indeed, multiplying the both sides of \re{em} by $x_j^{N-l}$ and taking
into account that according to the definition \re{VdM},
$\sum_j W_{kj} x_j^{N-l}=\delta_{jl}$, we get
\be
d \tau_k = \sum_{j=1}^{N-1}
\frac{dx_j\,x_j^{N-k}}{\sqrt{\Lambda^2(x_j)-4x_j^{2N}}}\,,
\lab{dtau}
\ee
where $k=2$, $...$, $N$. Comparing \re{dtau} with the definition \re{diff}
of the differentials $d\omega_k$ and $\Omega$ we obtain
\be
t_2\equiv \tau_2\,T = \sum_{j=1}^{N-1} \int^{Q_j} d\Omega\,,\qquad
t_k\equiv \tau_k\,T^{k-1} = \sum_{j=1}^{N-1} \int^{Q_j} d\omega_{k-2}\,,
\lab{tau}
\ee
where $k=3$, $...$, $N$ and the points $Q_j$ lie on the Riemann surface 
one each over the $\alpha-$cycles.
Finally, we introduce a new set of the Reggeon coordinates
\be
(Q_1,...,Q_{N-1})\to (\vartheta,\varphi_1,...,\varphi_{N-2})\,,
\lab{phi}
\ee
where
\be
\vartheta = -i \sum_{j=1}^{N-1} \int^{Q_j} d\hat\Omega\,,\qquad
\varphi_k = \sum_{j=1}^{N-1} \int^{Q_j} d\hat\omega_k
\lab{phi1}
\ee
and $k=1$, $...$, $N-2$.
Using the definitions \re{tau}, \re{bas-1} and \re{bas-3}
we find that the equations of motion \re{dtau} are trivially integrated
in these variables
\be
\vartheta = 2 t_2 + \sum_{j=1}^{N-2} V_j(u) t_{j+2}
\,,\qquad
\varphi_k = \sum_{j=1}^{N-2} U_{kj}(u) t_{j+2}\,,
\lab{wind}
\ee
with $U_{kj}$ and $V_j$ being $t-$independent functions of the moduli.

The following remarks are in order. The number of points $Q_j$ in \re{phi}
is equal to the number of allowed bands, $N-1=g+1$, and it does not match
the genus of the Riemann surface $\Gamma_N$. The vector with the
coordinates $\varphi=(\varphi_1,...,\varphi_{N-2})$ depends on the
integration path entering into \re{phi1} and it defines the point on the
Jacobian $\CJ(\Gamma_N)$ of the complex curve \re{GamN}. At the same time,
the variable $\vartheta$ is related to the differential $d\hat\Omega$ and
it starts to play a special role. Then, in new coordinates,
$\vartheta$ and
$(\varphi_1,...,\varphi_{N-2})\in \CJ(\Gamma_N)$, the relations \re{wind}
describe a ``fast'' winding of the Reggeons around the Jacobi torus of
the Riemann surface $\Gamma_N$ and a ``slow'' periodic motion in $\vartheta$.
The corresponding periods can be evaluated using \re{phi1}, \re{bas-1} and
\re{bas-3} as
$$
\oint_{\alpha_j}d\varphi_k= 2\pi \delta_{jk}\,,\qquad
\oint_{\alpha_{N-1}}d\varphi_k
=-\sum_{j=1}^{N-2} \oint_{\alpha_j}\varphi_k=-2\pi
$$
and
$$
\oint_{\alpha_j}d\vartheta=0\,,\qquad
\oint_{\alpha_{N-1}}d\vartheta=i\sum_{j=1}^{N-2} \oint_{\alpha_j} d\hat\Omega
-2\pi\res_{P_+}d\hat\Omega=-2\pi
\,,
$$
where $j,\,k=1$, $...$, $N-2$. We recall that the generators of 
``fast'' motion in times $t_3$, $...$, $t_N$ are the hamiltonians
$q_3$, $...$, $q_N$, while the hamiltonian $q_2$ generates ``slow''
drift of the system in time $t_2$.

\subsection{Action-angle variables}

The periods $a_j$ and the coordinates $\varphi_j$ have a simple interpretation
in terms of the action-angle variables for the $N$ Reggeon state.
The differential $dS_0$ becomes the generating function of the canonical
transformation from the separated coordinates $(x_j,p_j)$ to the
action-angle variables $(\phi_j,J_j)$. Namely, the action variables are
defined as
$$
J_k = \frac1{2\pi} \oint_{\alpha_k} dx \, p(x)
    = \frac1{2\pi} \oint_{\alpha_k} d S_0
\equiv \frac{a_k}{2\pi}
$$
and the corresponding angles are given by
$$
\phi_k=\frac{\partial S}{\partial J_k}\,,\qquad
S=\sum_{k=1}^{N-1} \int^{x_k} dx \, p(x) = \sum_{k=1}^{N-1} \int^{Q_k} dS_0\,,
$$
where $x_k$ belongs to the $k-$th allowed band. Substituting the
canonical form \re{can} of the differential $dS_0$ into this relation
and taking into account that $T=\sum_{j=1}^{N-1} J_k$
due to \re{sum}, one finds
\be
\phi_k=\varphi_k-\vartheta\,,\quad
\phi_{N-1}=-\vartheta\,,\quad
\oint_{\alpha_j} d\phi_{j'}=2\pi \delta_{jj'}\,,
\lab{varphi}
\ee
where $j,j'=1$, $...$, $N-1$ and the last identity follows from
\re{bas-1} and \re{bas-3}.
The angles $\phi_k$ describe the winding of the Reggeons around
$\alpha-$cycles on the Riemann surface $\Gamma_N$ and
the corresponding basic oscillation frequencies are defined as
$$
\phi_k = \Theta_{kj} \tau_j\,.
$$
The frequencies $\Theta_{kj}$ do not depend on the times $\tau_j$ and
they can be easily evaluated from \re{varphi} and \re{wind} in terms
of  the coefficients $U$ and
$V$ entering into definition of normalized differentials \re{bas-1}
and \re{bas-3}
\footnote{
The same expression can be found using the fact that the transition from
separated coordinates $(x_j,p_j)$ to the action-angle variables
$(J_k,\varphi_k)$ is the canonical transformation and the evolution of
$\varphi$ is described by the Hamiltonian equation
$$
\frac{\partial\varphi_k}{\partial\tau_j}=\{\varphi_k,\hat q_j\}=
\frac{\partial\hat q_j}{\partial J_k}
=2\pi\frac{\partial\hat q_j}{\partial a_k}=
\Theta_{kj}\,,
$$
where periods were defined in \re{per} and \re{ak}.}.

Summarizing our consideration of the dynamics of the $N$ Reggeon system
governed by the leading term in the asymptotic expansion of the Baxter
equation solution \re{f}, we found that as a result of the composition
of the maps
$$
(z_j,\pi_j) \to (x_j,p_j) \to (Q_j) \in \Gamma_N \to
\vartheta,(\varphi_1, ..., \varphi_{N-2})\in \CJ(\Gamma_N)
$$
the Reggeons have linear trajectories \re{wind} on the Jacobian of the
Riemann surface $\Gamma_N$. Performing inverse transformations one
can construct the explicit expressions for the holomorphic Reggeon
coordinates by means of the Riemann theta functions corresponding
to $\Gamma_N$ \ci{D,Kr}. The resulting expressions are similar to the
soliton solutions of the KP/Toda hierarchy \ci{Kr,NMPZ,D} and in the
center-of-mass frame of the $N$ Reggeon compound state, $P=0$, they can be
represented as
\be
z_k = z_k(t_2,t_3,...,t_N)
=\exp\lr{2i t_2}\cdot
\Psi_k(t_3 U^{(1)} + ... + t_N U^{(N-2)};u_1,...,u_{N-2})\,.
\lab{soliton}
\ee
Here, $\Psi_k(\varphi_1,...,\varphi_{N-2}; u_1,...,u_{N-2})$ is a
$2\pi$ periodic function of the variables $\varphi_j$ given by \re{wind}.
It depends on the parameters
$u_k$ defined in \re{hk} and on the $(N-2)-$dimensional wave vectors
$U^{(k)}=(U_{1k},...,U_{N-2,k})$ built from the matrix $U$. In
analogy with the KP/Toda solitons \ci{Kr,D,NMPZ}, the wave vectors
can be expressed in terms of the $\beta-$periods of the
differentials of the second kind, $d\hat \Omega_k$, on the curve
$\Gamma_N$, normalized as \ci{D}
$$
\oint_{\alpha_j}d\hat \Omega_k=0\,,\qquad
d\hat \Omega_k \stackrel{\xi\to 0}{\sim} \frac{d\xi}{\xi^{k+1}}\,,
$$
where $\xi=T/x\to 0$ is a local parameter on the curve $\Gamma_N$ at the
vicinity of infinity $P_+$ on the upper sheet. Using the well-known
property \ci{D}, that the $\beta-$periods of the differentials $d\hat 
\Omega_k$ are related to the behaviour of the holomorphic differential
$d\hat\omega_k$ near infinity $P_+$ as
$$
\oint_{\beta_j}d\hat
\Omega_k=\frac1{k!}
\partial_\xi^{k-1}
f_j(\xi)\left.\right|_{\xi=0}\,,\qquad
d\hat\omega_j \stackrel{\xi\to 0}{\sim}  f_j(\xi) d\xi
$$
and substituting \re{bas-1} and \re{diff} into this
relation, one can obtain the following expressions for the
coefficients $U_{kj}\equiv U^{(j)}_k$
\be
U^{(1)}_k = 2i\oint_{\beta_k}d\hat\Omega_1\,,\qquad
U^{(2)}_k+ \frac{u_1}2 U^{(1)}_k =
4i\oint_{\beta_k}d\hat\Omega_2\,,\ \
... \ \ .
\lab{beta}
\ee

To understand the origin of the first factor in \re{soliton} involving
the time $t_2=\tau_2 T$, we observe that \re{soliton} gives an exact
solution to the hierarchy of the conservation laws
$$
\frac{\partial z_k}{\partial \tau_j}=\{ z_k,\hat q_j \}\,.
$$
Let us consider the evolution of $z_k$ in time $\tau_2$. Replacing in
\re{q2} $\pi_j=-i\eta\partial_{z_j}$ by their classical analogs,
one finds that the Hamiltonian $\hat q_2$ is related to the holomorphic part
of the total angular momentum of the $N$ Reggeon system in the center-of-mass
rest frame, $\sum_j\pi_j=P=0$, as
$$
\hat q_2\equiv -T^2 = \lr{\sum_{k=1}^N z_k \pi_k}^2\,.
$$
Therefore, the evolution of the Reggeon holomorphic coordinates
in time $\tau_2$ becomes trivial
$$
\partial_{\tau_2} z_k = \{z_k,\hat q_2\} = 2iT z_k
\Longrightarrow z_k \sim \exp\lr{2i \tau_2 T}
$$
and it corresponds to the rotation of the $N$ Reggeon system on
the 2-dimensional plane of the impact parameters \re{imp} with the
angular momentum $2T$. Then, using the definitions \re{T1}, \re{q2}
and \re{h-q} one obtains that in the large $h$ limit the total angular
momentum of the soliton wave is equal by the Lorentz spin $m$ of
the $N$ Reggeon state
$$
2T = 2\sqrt{h(h-1)}= 2\lr{h-\frac12}+\CO(h^{-1})=m\,.
$$
Thus, the first factor in \re{soliton} describes the rotation of the
$N$ Reggeon system around its center-of-mass in the 2-dimensional
impact parameter space.  For $N=2$ Reggeon state, the BFKL Pomeron, this
becomes the only mode of the Reggeon motion. For $N\ge 3$ the quasiclassical
dynamics of the $N$ Reggeon compound states becomes much more interesting
due to appearance of the soliton excitations \re{soliton}. The space of the
parameters $u_1$, $...$, $u_{N-2}$ of the $N$ Reggeon soliton waves
coincides with the moduli space $\CM$ of the hyperelliptic curve $\Gamma_N$.
The coordinates on $\CM$ are determined by the values of the quantum numbers
$q_2$, $...$, $q_N$.

\subsection{Singularities on the moduli space}

The numerical solutions of the Baxter equation \ci{Bet,Qua} indicate that
for fixed $q_2$ the possible values of the quantum numbers $q_3$, $...$,
$q_N$ occupy a compact region in $(N-2)-$dimensional space with the boundary
defined by the hypersurface \re{crit}. Let us show that the hypersurface
\re{crit} can be identified from the analysis of singularities on the
moduli space $\CM$ of the $N$ Reggeon soliton waves. It is well-known
\ci{D,NMPZ} that the curve $\Gamma_N$ becomes singular only when two branch
points merge, $\sigma_j=\sigma_{j+1}$ (see Fig.~1). Since the positions
of $\sigma_j$ depend on the values of the quantum numbers, the latter
relation implies certain conditions on $q_k$.

For $N=2$ Reggeon states, one finds the branch points of the
curve \re{Gamma2} as
$$
-\sigma_1=\sigma_2=\frac{T}2\,.
$$
Since $T^2=-\hat q_2 > 0$ for the polynomial solutions, the values of
$\sigma_1$ and $\sigma_2$ are always real and different%
\footnote{We recall that the band $[\sigma_1,\sigma_2]$ defines the
interval of classical motion in the separated coordinate $x_1$.}.
Thus, the curve $\Gamma_2$ does not have any singularities and
this is in agreement with the fact \ci{FK,Qua} that the exact solution 
of the Baxter equation for $N=2$ is well defined and the asymptotic
expansion of the energy \re{as-exp} is Borel summable for $N=2$.

For $N=3$ Reggeon states, or the QCD Odderon, the curve \re{Gamma3}
has four branching points%
\footnote{Here we do not require the ordering
$\sigma_1 \le \sigma_2 \le \sigma_3 \le \sigma_4$.}
$$
4\sigma_j^3+\hat q_2\sigma_j+\hat q_3=0\,, \quad (j=1,2,3)\,,
\qquad
\sigma_4=-\frac{\hat q_3}{\hat q_2}\,.
$$
To classify all possible solutions we introduce the ``effective''
discriminant $\Delta$ of the curve $\Gamma_3$ as
$$
\Delta=16 \sigma^2_4
       (\sigma_1-\sigma_2)^2(\sigma_2-\sigma_3)^2(\sigma_3-\sigma_1)^2
      = \hat q_2^4\ u_1^2(1-27 u_1^2)\,,
$$
where $u_1$ was defined in \re{hk}. Then, for $\Delta > 0$
all roots $\sigma_k$ are real; for $\Delta < 0$ two roots
are real and two remaining roots are complex conjugated to each other
and, finally, for $\Delta=0$ two roots coincide and the
curve $\Gamma_3$ becomes singular.

Therefore, in order to be
able to construct two real intervals corresponding to the classical
motion of the Reggeons in the separated coordinates $x_1$ and $x_2$
one has to require $\Delta > 0$. If one assumes that
the moduli space $\CM$ is the complex $u_1-$plane, then the
singularities are located at three points
\be
u_1^{\rm crit}= -\frac1{\sqrt{27}} \,,\ 0 \ \ {\rm and} \ \ \frac1{\sqrt{27}}
\lab{h3-crit}
\ee
and the polynomial solutions of the Baxter equation for $N=3$ Reggeon
states correspond to the real values of $u_1$ shown in Fig.~2 such that
\be
0 < u_1^2 < \frac1{27}\,.
\lab{h3-int}
\ee
For $\lambda=-1$ the property \re{sym} of the curve $\Gamma_3$ leads
to the symmetry on the moduli space $\CM$ under
$u_1\leftrightarrow -u_1$.

\begin{figure}[hbt]
   \centerline{
   \epsfysize=6cm\epsfxsize=7cm
   \epsffile{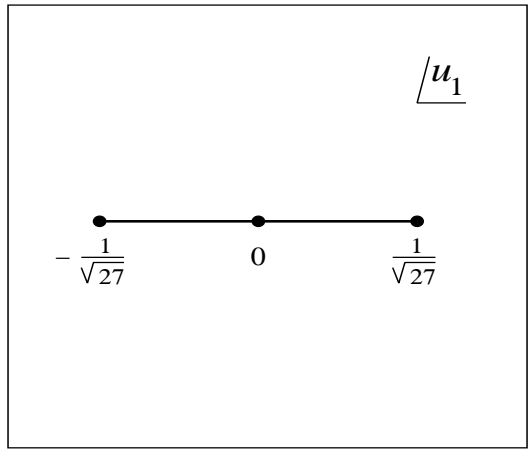}}
\caption{The possible values of $u_1$ for $N=3$ Reggeon states
corresponding to the polynomial solutions of the Baxter equation.
The dots indicate the positions of singularities.}
\end{figure}

The relations \re{h3-crit} and \re{h3-int} are in complete agreement
with the numerical results \ci{Bet,Qua} and with asymptotic expansions of the
solutions of the Baxter equation, \re{q3-cr}. A similar analysis
can be carried out for higher $N > 3$ Reggeon states \ci{Qua}.

Any point on the moduli space $\CM$ corresponds to a certain Riemann
surface $\Gamma_N$ and it describes the compound state of
$N$ Reggeons with the quantum numbers defined by \re{T1}. The energy
$\varepsilon_N$ of these states becomes the function on the moduli space
and it is naturally to expect that singularities on the moduli space
$\CM$ control the analytical properties of the functions $\varepsilon_N$
for $N\ge 3$. They are responsible for the appearance of Borel
singularities in the asymptotic expansion of the energy \re{as-exp}
for $N\ge 3$.

\sect{Quantization conditions}

Let us discuss the quantization conditions for eigenvalues
of the integrals of motion $q_2$, $...$, $q_N$. The quantization of $q_2$
follows from the definition \re{q2} after one takes into account that 
possible values of the conformal weight are given by \re{h-q}.
The polynomial solutions of the Baxter equation correspond to the special
values \re{h+} of quantized $q_2=-h(h-1)$. In this case one has to establish
the quantization conditions for the remaining charges $q_3$, $...$, $q_N$
and then try to generalize them from integer conformal weight $h$ to
all possible complex values \re{h-q}.

Inverting the dependence \re{ak} and using \re{sum} one can obtain that
the quantized values of the moduli \re{hk}, or equivalently
$q_3$, $...$, $q_N$, are determined by $q_2$ and by $N-2$
linear independent periods $a_k$ of the differential $dS_0$
defined in \re{per}
\be
u_k=u_k(\hat q_2; a_1,a_2,...,a_{N-2})\,,\qquad k=1,...,N-2\,.
\lab{hk-q}
\ee
Although $\hat q_2$ and $a_k$ depend separately on the parameter
$\eta$, this dependence is cancelled inside the function $u_k$
due to invariance of the moduli under transformation \re{sym}.

The periods $a_k$ have a simple interpretation in terms of the roots
$\lambda_j$ of the solutions \re{poly} of the Baxter equation. We recall
that the solutions $Q(x)$ define the wave function of the $N$ Reggeon
state in the separated coordinates \re{wf}. Therefore, the roots $\lambda_j$
of $Q(x)$ being the zeros of the Reggeon wave function should belong
to the intervals of the classical motion of Reggeons, that is to the
$N-1$ allowed bands on the Riemann surface $\Gamma_N$. Let us denote
by $n_k$ the number of roots of $Q(x)$ (including the
$N-$time degenerate root at $x=0$) which belong to the $k-$th allowed band,
$\lambda_j \in [\sigma_{2k-1},\sigma_{2k}]$,
\be
n_k \ge 0\,,\qquad \sum_{k=1}^{N-1} n_k = h
\lab{sets}
\ee
and $n_j \ge N$ if the root $x=0$ belongs the $j-$interval
$\sigma_{2j-1} < 0 < \sigma_{2j}$.
Then, it follows from \re{S} that the meromorphic differential $dS$ has
first-order poles on the Riemann surface $\Gamma_N$ at $x=\lambda_j$
and its periods around $\alpha-$cycles count the number of roots
$$
\oint_{\alpha_k} dS = 2\pi \eta n_k\,.
$$
Substituting the expansion \re{exp} of $dS$ in powers of $\eta$
one gets
\be
a_k\equiv \oint_{\alpha_k} dS_0 = 2\pi \eta\lr{n_k+\frac12}
                                + \CO(\eta^2)\,,
\lab{BS}
\ee
where $\CO(\eta^2)$ term takes into account the contribution of
higher terms in the expansion \re{exp}. The relations \re{BS} take the form
of Bohr-Sommerfeld quantization conditions for the Reggeon wave
function \re{f}.

Substituting \re{BS} into \re{hk-q} we find that the moduli of
the Riemann surface, or equivalently the integrals of motion $q_3$, $...$,
$q_N$, become quantized and, in accordance with our expectations \re{int},
their values are parameterized by $q_2$ and by the set of positive
integer numbers $n_k$. However, trying to find the dependence of moduli
on $n_k$ from \re{hk-q} and \re{BS}, one has to take into account that
the periods $a_k$ themselves are functions of the moduli due to
$\CO(\eta^2)$ term in \re{BS}, which does not depend on integers
$n_k$ and has the general form $\eta^2\lr{-\hat q_2}^{1/2} H(\{u_k\})$.
Here, $H$ is a complicated function of the moduli and $\hat q_2$ is needed
to restore the scaling dimension of $a_k$ in $\eta$ under transformation
\re{sym}. Then, using independence of $u_k$ on $\eta$ one can put
$\eta=\lr{-q_2}^{-1/2}$ in \re{hk-q} and represent the
general solution of \re{hk-q} and \re{BS} as
\be
u_k=\bar u_k\lr{\delta_1,\delta_2,...,\delta_{N-2};q_2^{-1}}\,,
\qquad \delta_j\equiv\frac{n_j+\frac12}{\lr{-q_2}^{1/2}}\,.
\lab{hbar}
\ee
One possibility to define the function $\bar u_k$ is to find its
asymptotic expansion in inverse powers of the conformal weight using
the series \re{as-exp},
\be
u_k=\sum_{k\ge 0} c_k(n)\, h^{-k}
\lab{ori}
\ee
with $q_2=-h(h-1)$ and $n_j=\mbox{fixed}$. However, due to the presence of
singularities on the moduli space $\CM$, the series  \re{ori} turns out to be
non Borel summable. Let us now change the parameters of
the expansion and expand the function \re{hbar} in powers of $q_2^{-1}$
keeping $\delta_j=\mbox{fixed}$. As example, one uses the result \ci{Qua} for
the large $h$ expansion of $u_1$ to $\CO(h^{-8})$ order for $N=3$ Reggeon
states to convert it into the following form
\be
u_1= \sum_{k=0}^\infty u_1^{(k)}(\delta)\,q_2^{-k}\,,
\lab{hq-exp}
\ee
where the leading term is given by
\be
u_1^{(0)}(\delta)=\frac1{\sqrt{27}}\left (1-3\,\delta+2\,{
\delta}^{2}-{\frac {2}{9}}{\delta}^
{3}+{\frac {10}{81}}{\delta}^{4}+{
\frac {38}{243}}{\delta}^{5}+{
\frac {448}{2187}}{\delta}^{6}+{
\frac {1840}{6561}}{\delta}^{7}
+\CO(\delta^8)
\right )\,.
\lab{fun}
\ee
We notice that the expansion \re{hq-exp} goes over integer powers of
$1/q_2=\CO(h^{-2})$, while the original series \re{ori} had a much
bigger parameter of the expansion, $1/h$.

Let us consider now the asymptotic approximation to solution of the
Baxter equation given by the expression \re{f}, in which we neglect all
nonleading $\CO(\eta)$ corrections to the exponent. In this limit,
there are no $\CO(\eta^2)$ corrections to the periods $a_k$ in
\re{BS} and the expressions \re{hk-q} and \re{hbar} for quantized moduli
look like
\be
u_k^{(0)}=u_k(\hat q_2;a_1,a_2,...,a_{N-2})
\big|_{a_k=2\pi\eta\lr{n_k+\frac12}}
=\bar u_k(\delta_1,\delta_2,...,\delta_{N-2};0)\,,
\lab{h-as}
\ee
where the parameters $\delta_k$ were defined in \re{hbar}.

To obtain all possible values of quantum numbers $q_3$, $...$, $q_N$ one
has to evaluate the moduli \re{h-as} for different sets of integers
$n_1$, $...$, $n_{N-2}$ and substitute them into \re{T1}. Let us
consider \re{h-as} as a definition of a continuous function of real
positive $q_2$, which for $q_2=-h(h-1)$ and $h$ positive integer gives
quantized $q_k$. Then, for different possible sets \re{sets}
of integers $n_k$ the functions \re{h-as} define the family of curves
on the moduli space $\CM$. Each curve describes the flow on
the moduli space $\CM$ in the ``slow'' time $q_2$ and it has a
distinguished property that the values of the periods $a_k$ are
preserved. As example \ci{Qua}, the flow of quantized $q_3$ as a
function of $q_2$ for $N=3$ Reggeon states and fixed value of integer
$n_1=4$ is shown by solid line in Fig.~3. This curve induces the flow
on the moduli space indicated by the solid line in Fig.~2.

\begin{figure}[hbt]
\vspace*{-10mm}
   \centerline{
   \epsfysize=10cm\epsfxsize=11cm
   \epsffile{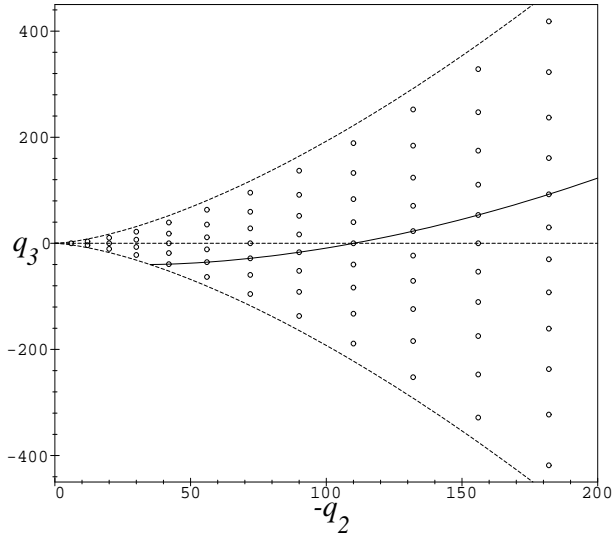}}
\vspace*{-10mm}
\caption{
One of the curves describing the flow of quantized $q_3$ for $N=3$
Reggeon states. The circles represent the results of numerical
calculations. Three dashed lines correspond to the singularities
on the moduli space \re{h3-crit}.
}
\end{figure}

For $N=3$ Reggeon states, $u_1^{(0)}$ is equal to the leading
term of the expansion \re{hq-exp} given by \re{fun}.
Comparing \re{fun} with \re{h3-crit} we find that the expression \re{fun}
provides a weak coupling expansion of the moduli in $\delta$
around one of the singularities on the moduli space in Fig.~2.
To approach two remaining singular points one has to develop
the strong coupling expansion of the moduli.

The Seiberg-Witten formalism \ci{SW} gives us a powerful method of
calculating the moduli \re{h-as}, based on the Whitham equations and on
the remarkable property of duality between strong and weak coupling
expansions of the moduli $u_k$ in parameters $\delta_j$ \ci{pre}.
In application to the $N=3$ Reggeon states, the duality originates
from the property \ci{SW} that the quantity $(a_1(u_1),a^D_1(u_1))$,
built from the periods \re{per} of the differential $dS_0$ on the curve
$\Gamma_3$, has well-known monodromies around the singular points
\re{h3-crit}. The monodromies are given by $2\times 2$ matrices, which
belong to the subgroup $\Gamma(2)$ of the $SL(2,\IZ)$ group consisting
of the matrices congruent to 1 modulo 2.%
\footnote{This explains the observation
made in \ci{J} that the Schr\"odinger equation for the $N=3$ Reggeon
state, QCD Odderon, obeys a new modular symmetry with respect to
$\Gamma(2)$.}
The symmetry of the spectral curve $\Gamma_3$ under $u_1\to -u_1$ leads
to the following property of the moduli \re{fun} \ci{pre}
$$
u_1(1-\delta)=-u_1(\delta)\,,
$$
which together with \re{fun} allows us to identify the values of
$\delta$ corresponding to the singularities \re{h3-crit} on the moduli
space as
$$
\delta^{\rm crit}= 1 \,,\ \frac12 \ \ {\rm and} \ \ 0\,,
$$
respectively. Having expressions for the monodromy of
$(a_1(u_1),a^D_1(u_1))$ one can determine the asymptotic behaviour of
the moduli around these points in the following form \ci{pre}
\be
\delta - 1 \sim -\sqrt{3}\lr{u_1+\frac1{\sqrt{27}}}
\,,\qquad
\delta-\frac12 \sim \frac3{\pi} u_1\lr{\ln u_1-1}
\,,\qquad
\delta \sim -\sqrt{3}\lr{u_1-\frac1{\sqrt{27}}}\,.
\lab{asymp}
\ee
These relations describe the flow of the quantum numbers shown in Fig.~3
and they can be also derived from the Whitham equations.

\subsection{Whitham equations}

Let us show that the flow of the moduli \re{h-as} in the ``slow'' time --
variable $q_2$ is governed by the Whitham equations \ci{Wh,IM}. We recall that
the function \re{hk-q} is inverse to \re{ak} and the dependence of $u_k$ on
$q_2$ in \re{h-as} can be found from the condition that the periods $a_k$
should be invariant under variations of $q_2$. Using the definition of the
periods \re{per}, this condition can be expressed as follows
\be
\delta a_k = \delta T \oint_{\alpha_k} \partial_{T} dS_0
           + \delta u_j \oint_{\alpha_k} \partial_{u_j} dS_0
           = 0\,.
\lab{var}
\ee
To calculate the external derivatives one considers the variation of
the differential $dS_0$ with respect to Reggeon quantum numbers.
As follows from the definitions \re{dS0} and \re{y}
$$
\delta dS_0 = dx \frac{\delta \omega}{\omega}
=\delta \Lambda(x) \frac{dx}{y}
=\lr{\delta q_2 x^{N-2} + \delta q_3 x^{N-3} + ... + \delta q_N}\frac{dx}{y}
\,,
$$
where we put $\eta=1$ for simplicity. Finally, one gets
$$
\frac{\partial dS_0}{\partial u_j} =  T d \omega_j\,,\qquad
\frac{\partial dS_0}{\partial T} =  T^{-1} d S_0 \,.
$$
where holomorphic differentials $d\omega_k$ were defined in \re{diff}.
Here, the first relation states that the variation of the differential
$dS_0$ with respect to moduli is proportional to the holomorphic
differential defined on the curve $\Gamma_N$. Being applied to \re{var},
this remarkable property of the differential $dS_0$ leads to the Whitham
equations for the moduli
\be
\sum_{j=1}^{N-2}
\frac{\partial u_j}{\partial T} T^2 \oint_{\alpha_k} d\omega_j+
\oint_{\alpha_k} dS_0=0 \,.
\lab{Wh-eq}
\ee
Calculating the $\alpha-$periods of the differential $d\omega_k$ from
\re{bas-1} as $\oint_{\alpha_k}d\omega_j=2\pi(U^{-1})_{jk}$ and taking
into account \re{BS}, we obtain the following system of equations
\be
T^2\frac{\partial u_j}{\partial T}=-\sum_{k=1}^{N-2}
\lr{n_k+\frac12}U_{kj}(u)\,,
\lab{Wh}
\ee
where $j=1$, $...$, $N-2$ and $U_{kj}$ are functions of the moduli
$u_1$, $...$, $u_{N-2}$. The matrix elements of $U_{kj}$ define
the wave vectors of the soliton waves \re{soliton} and one can rewrite
\re{Wh} as
$$
T^2\frac{\partial u_j}{\partial T}=- (v, U^{(j)})\,,
$$
where notation was introduced for the vector $v_k=n_k+\frac12$
and $(v,U^{(j)})\equiv v_k U_{kj}$.

For $N=3$ Reggeon states, the system \re{Wh} is reduced to an
ordinary differential equation for the moduli $u_1=u_1(\delta)$
with $\delta=(n+\frac12)/T$
\be
\frac{d u_1}{d \delta}= U(u_1)\,,
\lab{W-3}
\ee
where $U=U^{(1)}$ was defined in \re{beta}.
It allows us to determine {\it exactly\/} the function $u_1^{(0)}$
entering into \re{fun} and study its properties at the vicinity of
singularities \re{crit} on the moduli space $\CM$. One can show \ci{pre}
that the Whitham equation \re{W-3} is in agreement with the
weak coupling expansion \re{fun} and with the asymptotics \re{asymp}.

\subsection{Boundary conditions}

To solve the differential equations \re{Wh} and \re{W-3} one has to
supplement them by an appropriate boundary conditions on $u_k$. These
conditions are provided by the asymptotic behaviour of $u_k$ as
$q_2\to-\infty$, which follows from the large $h$ expansion of quantum
numbers \re{as-exp}.

Let us start with $N=3$ Reggeon states and examine the flow of quantized
$q_3$ shown by solid line in Fig.~3. For $-q_2=h(h-1)\to \infty$
the quantized $q_3$ behave as $q_3 \sim (-q_2)^{3/2}/\sqrt{27}$
leading to the asymptotic expression for the moduli $u_1=\frac1{\sqrt{27}}$,
which we identify as one of the singularities on the
moduli space $\CM$. Thus, the singularity on the moduli space,
\be
u_1(\delta)=\frac1{\sqrt{27}}\,, \qquad \mbox{for} \quad
\delta=0
\lab{boun}
\ee
becomes the starting point for the Whitham evolution \re{W-3}.
As $-q_2$ decreases toward the origin, $u_1$ passes the second
singularity \re{crit} at $u_1=0$, and for smallest $-q_2$, corresponding
to the boundary of the polynomial solutions, $u_1$ approaches the
third singularity on the moduli space at $u_1=-\frac1{\sqrt{27}}$.
We conclude that for polynomial solutions of the $N=3$ Baxter equation,
the Whitham equation \re{W-3} describe the trajectory on the moduli
space (see Fig.~2), which goes along the real axis from $\frac1{\sqrt{27}}$
to $-\frac1{\sqrt{27}}$.
The Whitham evolution of the moduli corresponding to higher
$N > 3$ Reggeon states follows a similar pattern.
For $-q_2\to\infty$ the quantum numbers $q_k$ approach the values
belonging to the critical hypersurface \re{crit}. The same values coincide
with the positions of singularities on the moduli space $\CM$,
$u_k\sim u_k^{\rm sing}$ as $-q_2\to\infty$.
Since the possible values of the moduli corresponding to the
polynomial solutions occupy a compact region on the moduli
space with the boundary being the singularities, the flow
of $u_k$ will start at one singular point as $-q_2\to\infty$ and
will finish at another singular point on $\CM$ for $-q_2$
taking the smallest value allowed for the polynomial solutions
of the Baxter equation.

As was shown in Sect.~4.2, the ``time'' variable in the Whitham evolution,
T, is related to the total angular momentum associated with the
rotation of the $N$ Reggeon system around its center-of-mass in the
impact parameter space. The Whitham equations \re{Wh} and \re{W-3}
describe the adiabatic perturbation of the moduli of the curve
$\Gamma_N$ which enter as parameters into the soliton solutions
\re{soliton} for the $N$ Reggeon states. Thus, in the leading
nonlinear WKB approximation, \re{f}, the $N$ Reggeon states corresponding
to the integer positive conformal weight $h$ can be considered
as modulated soliton waves.

We recall that till now we worked on the subspace of the polynomial
solutions of the Baxter equation. The Whitham equations \re{Wh} and \re{W-3}
offer a natural way of analytical continuation of the obtained expressions
beyond this subclass. Let us consider for simplicity the $N=3$ Reggeon
states. The evolution of $q_3$ is described by a smooth function of
$q_2=-h(h-1)$ (see Fig.~3), which for $h$ in \re{h+} gives
the values of quantized $q_3$. We observe that, first, the same function
allows us to define formally $q_3$ corresponding to half-integer $h$
in \re{h-q} and, second, the evolution of $q_3$ inevitably leads to the
region of small and positive $q_2$, where polynomial solutions do not exist.
Applying the Whitham equations \re{Wh} and \re{W-3} in these two cases one
assumes that the above interpretation of the $N$ Reggeon states as
modulated soliton waves holds for any integer and half-integer conformal
weight $h$. One may also try to apply the Whitham equations for an arbitrary
complex quantized conformal weights \re{h-q}, but the important difference
with real $h$ is that one has to find now the initial conditions
similar to \re{boun}.

\sect{Conclusions}

The Regge asymptotics of hadronic scattering amplitudes in high-energy
QCD are controlled by the color-singlet compound states of Reggeons.
Reggeons appear as a new collective degrees of freedom of QCD
in the Regge limit and their dynamics in four dimensions is described by
the effective (1+1)-dimensional Hamiltonian, which exhibits remarkable
properties of integrability. The system of $N$ interacting Reggeons
in the multi-color limit resembles very much the 1-dimensional Heisenberg
spin $s=0$ chain with $N$ sites. It has enough number of conserved
charges $q_k$ to be completely integrable. To diagonalize the
$N$ Reggeon hamiltonian and calculate the spectrum of the Reggeon
compound states, QCD Pomerons and Odderons, we defined a new set of
Reggeon coordinates, in which $N$ coupled Schr\"odinger equations
for eigenvalues of the conserved charges $q_k$ become separated and are
replaced
by the Baxter equation \re{Bax}. The solutions of the Baxter equation,
depending on the set of quantum numbers $q_k$, define the energy and
the wave function of the $N$ Reggeon state in the separated variables.
It is the Baxter equation that summarizes the QCD dynamics of
Reggeons in the separated coordinates and whose nonlinear WKB expansion
gives rise to the integrable structures well-known from the finite-gap
solutions of the soliton equations and their Whitham deformations.

The leading term of the nonlinear WKB expansion of the polynomial
solutions of the Baxter equation defines the hyperelliptic Riemann
surface $\Gamma_N$ as the level surface of the integrals of motion
$q_k=\mbox{const.}$ and the meromorphic 1-differential $dS_0$ on
it. The conserved charges $q_k$ generate the Hamiltonian flows of
Reggeons on $\Gamma_N$ and the exact solution of the arising
hierarchy of the evolution equations is given by the Reggeon soliton
wave. The moduli $u_k$ of $\Gamma_N$ for $N\ge 3$ depend on the quantum
numbers of the Reggeon states and enter as parameters into the soliton
solutions. The possible values of $q_3$, $...$, $q_N$ are quantized
and they determine the family of curves on the moduli space $\CM$.
Each curve describes the flow on $\CM$, which is governed by the
Whitham equations. These equations describe the adiabatic perturbation
of the Reggeon soliton waves and the properties of their solutions
will be considered in the forthcoming paper \ci{pre}.

We would like to mention in conclusion that had we perform a similar
nonlinear WKB analysis of the Baxter equation for Toda chain \re{Toda}
in the special case $\hbar=1$, we could reproduce the main ingredients
of the Seiberg-Witten solution \ci{SW} of the effective action
of $N=2$ supersymmetric QCD. It remains unclear, however, what are a 
new collective variables in $N=2$ SUSY QCD, which play a role 
similar to the separated coordinates of Reggeons in high-energy QCD.

\section*{Acknowledgements}

The author is most grateful to A.S. Gorsky, V.P. Spiridonov and
G. Veneziano for stimulating discussions.

\bb{99}
\bi{book} G. Veneziano, Phys. Rep. 9 (1974) 199; 
\\        A.M. Polyakov, {\it Gauge fields and strings\/}, Chur,
          Harwood, 1987.
\bi{SW}   N. Seiberg and E. Witten,
          Nucl. Phys. B426 (1994) 19; B431 (1994) 484.
\bi{FK}   L.D. Faddeev and G.P. Korchemsky, preprint ITP--SB--94--14,
          Apr. 1994 [hep-ph/9404173]; Phys. Lett. B 342 (1995) 311.
\bi{Lip}  L.N. Lipatov, JETP Lett. 59 (1994) 596.
\bi{gen}  A. Klemm, W. Lerche, S. Yankielowicz and S. Thiesen,
          Phys. Lett. B344 (1995) 169;
\\        P.C. Argyres and A. Faraggi, Phys. Rev. Lett. 74 (1995) 3931;
\\        U.H. Danielsson and B. Sundborg, Phys. Lett. B358 (1995) 273;
\\        A. Brandhuber and K. Landsteiner, Phys. Lett. B358 (1995) 73;
\\        A. Hanany and Y. Oz, Nucl. Phys. B452 (1995) 283.
\bi{GKM3} A. Gorsky, I. Krichever, A. Marshakov, A. Mironov and A. Morozov,
          Phys. Lett. B355 (1995) 466.
\bi{To}   E. Martinec and N. Warner, Nucl. Phys. B459 (1996) 97;
\\        T. Nakatsu and K. Takasaki, Mod. Phys. Lett. A11 (1996) 157;
\\        T. Eguchi and S. Yang, preprint UT-728 [hep-th/9510183].
\bi{CM}   R. Donagi and E. Witten, Nucl. Phys. B460 (1996) 299;
\\        E. Martinec, Phys. Lett. B367 (1996) 91;
\\        A. Gorsky and A. Marshakov,
          preprint FIAN-TD-19-95, Oct 1995 [hep-th/9510224];
\\        E. Martinec and N. Warner, preprint EFI-95-70, Nov. 1995
          [hep-th/9511052].
\bi{IM}   H. Itoyama and A. Morozov, preprints ITEP-M5/95, Nov. 1995
          [hep-th/9511126] and ITEP-M6/95, Dec. 1996 [hep-th/9512161].
\bi{spin} A. Gorsky, A. Marshakov, A. Mironov and A. Morozov,
          Phys. Lett. B380 (1996) 75; preprint ITEP/TH-9/96, Apr. 1996
          [hep-th/9604078].
\bi{Wh}   G.B. Whitham, {\it Linear and Nonlinear Waves\/},
          John Wiley, New York, 1974;
\\        H. Flaschka, M.G. Forest and D.W. McLaughlin, Comm. Pure Appl.
          Math. 33 (1980) 739;
\\        S.Yu. Dobrokhotov and V.P. Maslov, J. Sov. Math. 16 (1981) 1433;
\\        B.A. Dubrovin and S.P. Novikov, Russ. Math. Surv. 44 (1989) 35.
\\        I.M. Krichever, Comm. Math. Phys. 143 (1992) 415;
          Comm. Pure Appl. Math. 47 (1994) 437;
\\        B.A. Dubrovin, Comm. Math. Phys. 145 (1992) 195.
\bi{TN}   K. Takasaki and T. Nakatsu, preprint KUCP--0092, Mar. 1996
          [hep-th/9603069].
\bi{G}    A. Gorsky, preprint ITEP/TH-14/96 [hep-th/9605135].
\bi{Col}  S.C. Frautschi, {\it Regge poles and S-matrix theory\/},
          New York, W.A. Benjamin, 1963;
\\        V. de Alfaro and T. Regge, {\it Potential scattering},
          Amsterdam, North-Holland, 1965;
\\        P.D.B. Collins, {\it An introduction to Regge theory
          and high energy physics\/}, Cambridge University Press, 1977.
\bi{CW}   H. Cheng and T.T. Wu, {\it Expanding Protons: Scattering at
          High Energies\/}, MIT Press, Cambridge, Massachusetts, 1987.
\bi{Gr}   V.N. Gribov, Sov. Phys. JETP 26 (1968) 414;
          Nucl. Phys. B106 (1976) 189.
\bi{Leff} R. Kirschner, L.N. Lipatov and L. Szymanowski,
          Nucl. Phys. B425 (1994) 579.
\bi{VV}   E. Verlinde and H. Verlinde, preprint
          PUPT--1319, Sept. 1993 [hep-th/9302104].
\bi{BKP}  J. Bartels, Nucl. Phys. B175 (1980) 365;
\\        J. Kwiecinski and M. Praszalowicz, Phys. Lett. B94 (180) 413.
\bi{QISM} L.A. Takhtajan and L.D. Faddeev, Russ. Math. Survey 34 (1979) 11;
\\        E.K. Sklyanin, L.A. Takhtajan and L.D. Faddeev,
          Theor. Math. Phys. 40 (1980) 688;
\\        V.E. Korepin, N.M. Bogoliubov and A.G. Izergin, {\it Quantum
          inverse scattering method and correlation functions\/},
          Cambridge Univ. Press, 1993.
\bi{Bet}  G.P. Korchemsky, Nucl. Phys. B443 (1995) 255.
\bi{FM}   H. Flaschka and D. McLaughlin, Progr. Theor. Phys.
          55 (1976) 438.
\bi{Guz}  M. Gutzwiller, Ann. Phys. 133 (1981) 304.
\bi{SoV}  E.K. Sklyanin, {\it The quantum Toda chain\/},
          Lecture Notes in Physics, vol.\ 226, Springer, 1985, pp.196--233;
          {\it Functional Bethe ansatz\/}, in ``Integrable
          and superintegrable systems'', ed.\ B.A. Kupershmidt, World
          Scientific, 1990, pp.8--33;
          Progr. Theor. Phys. Suppl. 118 (1995) 35 [solv-int/9504001].
\bi{MW}   Z. Maassarani and S. Wallon, J. Phys. A: Math. Gen. 28 (1995) 6423.
\bi{Qua}  G.P. Korchemsky, Nucl. Phys. B462 (1996) 333.
\bi{J}    R. Janik, Phys. Lett. B371 (1996) 293;
          Acta Phys. Polon. B27 (1996) 1275.
\bi{JW}   R. Janik and J. Wosiek, presentation pa02-059 at the
          ICHEP--96, Warsaw, 25--31 July 1996.
\bi{NMPZ} S.P. Novikov, S.V. Manakov, L.P. Pitaevskii and V.E. Zakharov,
          {\it Theory of Solitons: The Inverse Scattering Method\/},
          Consultants Bureau, New York, 1984;
\\        B. Dubrovin, I. Krichever and S. Novikov,
          {\it Integrable systems - I},
          Sovremennye problemy matematiki (VINITI), Dynamical systems - 4
          (1985) 179;
\\        B.A. Dubrovin, V.B. Matveev and S.P. Novikov,
          Russ. Math. Surv. 31 (1976) 59.
\bi{Kr}   I.M. Krichever, Russ. Math. Surv. 32 (1977) 185;
          Func. Anal. Appl. 14 (1980) 531; 11(1977) 12.
\bi{BFKL} E.A. Kuraev,  L.N. Lipatov and V.S. Fadin,
          Phys. Lett. B60 (1975) 50;
          Sov. Phys. JETP 44 (1976) 443; 45 (1977) 199;
\\        Ya.Ya. Balitsky and L.N. Lipatov, Sov. J. Nucl. Phys. 28 (1978) 822.
\bi{Lip2} L.N. Lipatov, Phys. Lett. B251 (1990) 284;  B309 (1993) 394.
\bi{Lip1} L.N. Lipatov, {\it Pomeron in quantum chromodynamic\/},
          in ``Perturbative QCD'', pp.411--489, ed. A.H. Mueller,
          World Scientific, Singapore, 1989.
\bi{XXX}  V.O. Tarasov, L.A. Takhtajan and L.D. Faddeev, Theor. Math.
          Phys. 57 (1983) 163.
\bi{R-m}  P.P. Kulish, N.Yu. Reshetikhin and E.K. Sklyanin, Lett. Math.
          Phys. 5 (1981) 393.
\bi{H}    M. Henon, Phys. Rev. B9 (1974) 1921;
\\        H. Flaschka, Phys. Rev. B9 (1974) 1924; Progr. Theor. Phys.
          52 (1974) 703;
\\        S.V. Manakov, Sov. Phys. JETP 40 (1974) 269.
\bi{Q}    R.J. Baxter, {\it Exactly Solved Models in Statistical
          Mechanics\/}, Academic Press, London, 1982;
          Stud. Appl. Math. 50 (1971) 51.
\bi{SZ}   V. Spiridonov, L. Vinet and A. Zhedanov, Lett. Math. Phys. 29
          (1993) 63;
\\        V. Spiridonov and A. Zhedanov, J. Phys. A: Math. Gen. 28
          (1995) L589.  
\bi{PG}   V. Pasquier and M. Gaudin, J. Phys. A: Math. Gen. 25 (1992) 5243.
\bi{P-I}  S.P. Novikov, Func. Anal. Appl. 24 (1990) 296;
\\        I.M. Krichever, ETH preprint, Z\"urich, June 1990;
\\        G. Moore, Comm. Math. Phys. 133 (1990) 261;
\\        F. Fucito, A. Gamba, M. Martinelli and O. Ragnisco,
          Int. J. Mod. Phys. B6 (1992) 2123.
\bi{odd}  P. Gauron and B. Nicolescu, Phys. Lett. B260 (1991) 40;
\\        P. Gauron, L. Lukaszuk and B. Nicolescu, Phys. Lett. B294
          (1992) 298;
\\        B. Nicolescu, Nucl. Phys. (Proc.Suppl.) 25B (1992) 142.
\bi{vM}   P. van Moerbeke, Invent. Math. 37 (1976) 45.
\bi{D}    B.A. Dubrovin, Russ. Math. Surv. 36 (1981) 11.
\bi{pre}  G.P. Korchemsky, in preparation.
\eb
\end{document}